\newcommand{\CLASSINPUToutersidemargin}{16.08mm}
\documentclass[conference]{IEEEtran}
\IEEEoverridecommandlockouts
\usepackage{listings}
\usepackage{caption}
\usepackage{xcolor}
\usepackage{bbding}
\usepackage[linesnumbered,ruled, noend]{algorithm2e}
\let\labelindent\relax
\usepackage{enumitem}
\usepackage[utf8]{inputenc}
\usepackage{multirow}
\usepackage{array}
\usepackage{booktabs}
\usepackage{verbatim} 
\usepackage{verbatimbox}
\usepackage[tight,footnotesize]{subfigure}
\usepackage{algorithmic}
\usepackage[cmex10]{amsmath}
\usepackage{amssymb}
\usepackage{graphics}
\usepackage[pdftex]{graphicx}
\usepackage[hyphens]{url}
\usepackage{flushend}

\graphicspath{{pictures/}} 
\ifCLASSINFOpdf
\graphicspath{{./pictures/}}
\DeclareGraphicsExtensions{.pdf,.jpeg,.png}
\fi

\definecolor{dkgreen}{rgb}{0,0.6,0}
\definecolor{gray}{rgb}{0.5,0.5,0.5}
\definecolor{mauve}{rgb}{0.58,0,0.82}
\definecolor{groovyblue}{HTML}{0000A0}
\definecolor{groovygreen}{HTML}{008000}
\definecolor{darkgray}{rgb}{.4,.4,.4}

\lstdefinelanguage{Groovy}[]{Java}{
	keywordstyle=\color{groovyblue}\bf,
	stringstyle=\color{mauve}\ttfamily,
	keywords=[3]{each, findAll, groupBy, collect, inject, eachWithIndex, subscribe, unsubscribe},
	morekeywords={def, as, in, use, TRIGGER, CHECK, RUN, run, match, satisfy, fetch, do, foreach, branch, fetch1},
	moredelim=[is][\textcolor{darkgray}]{\%\%}{\%\%},
	moredelim=[il][\textcolor{darkgray}]{��}
}

\newcommand{\tool}{\textsc{PFirewall}\xspace}

\begin{document}

\title{\tool: Semantics-Aware Customizable Data Flow Control for Home Automation Systems\\
\thanks{An earlier version of this paper was submitted to USENIX Security on November
15th, 2018. This version contains some minor modifications based on that
submission.}
}


\author{\IEEEauthorblockN{Haotian Chi\IEEEauthorrefmark{1}, Qiang Zeng\IEEEauthorrefmark{2}, Xiaojiang Du\IEEEauthorrefmark{1}, Lannan Luo\IEEEauthorrefmark{2}}
\IEEEauthorblockA{\IEEEauthorrefmark{1}Department of Computer and Information Sciences, Temple University, Philadelphia, PA 19122, USA \\
\IEEEauthorrefmark{2}Department of Computer Science and Engineering, University of South Carolina, Columbia, SC 29208, USA \\
Email: \{htchi, dux\}@temple.edu, \{zeng1, lluo\}@cse.sc.edu}}

\maketitle

\begin{abstract}
Emerging Internet of Thing (IoT) platforms provide a convenient solution for integrating 
heterogeneous IoT devices and deploying home automation applications. 
However, serious privacy threats arise as device data now flow out
to the IoT platforms, which may be subject to various attacks. 
We observe two privacy-unfriendly practices in emerging home automation systems: first,
the majority of data flowed to the platform are superfluous in the sense that 
they do not trigger any home automation; second, home owners currently have nearly \emph{zero}
control over their data. 

We present \tool, a customizable data-flow control system to enhance user privacy.
\tool analyzes the automation apps to extract their semantics, which are
automatically transformed into data-minimization policies; these policies only send minimized data flows to the platform for app execution, such that the ability of attackers to infer user privacy is significantly impaired. 
In addition, \tool provides capabilities and interfaces for users to define and enforce customizable policies based on individual privacy preferences. 
\tool adopts an elegant man-in-the-middle design, transparently executing data-minimization and user-defined policies to process raw data flows and mediating the processed data
between IoT devices and the platform (via the hub), without requiring modifications of the platform or IoT devices.
We implement \tool to work with two popular platforms: SmartThings and openHAB, and set up two real-world testbeds to evaluate its performance. 
The evaluation results show that \tool is very effective: it reduces 
IoT data sent to the platform by 97\% and enforces user-defined policies successfully.  
\end{abstract}



%
\IEEEpeerreviewmaketitle

\section{Introduction}
With the prosperity of Internet of Things (IoTs), smart systems 
(e.g., smart homes, factories, and hospitals)
have become realistic and are expanding with an ever-increasing speed~\cite{iot2018platforms}. 
IoT Platforms, such as SmartThings, Wink, openHAB, allow smart home users to connect heterogeneous IoT devices (e.g., sensors, actuators, appliances) to a platform-provided hub and to install applications on the platform to create automatic interactions among devices, i.e., home automation. 

As IoT device data flow to the platform, 
protecting user privacy becomes critical~\cite{zheng2018user, zeng2017end}. 
Existing work protects user privacy by resolving threats caused by malicios automation applications~\cite{celik18sensitive, bastys2018if, fernandes2016flowfence, tian2017smartauth} or handling attacks that eavesdrop IoT device traffic~\cite{acar2018peek, datta2018developer, apthorpe2017closing, apthorpe2017spying}.
Surprisingly, none investigates privacy protection at the platform architectural level,
even though the platform receives huge amounts of data from smart homes
and has full data access privileges. Indeed, it is baseless to assume 
the platform is secure and trustworthy. A platform
could be compromised by both inside attackers~\cite{bupadisgruntledemployee}
and remote attackers that exploit the vulnerabilities of its hub and cloud~\cite{17biggestdatabreach}.
Compared to clouds that have suffered many notorious attacks, an IoT platform 
has a much larger attack surface involving not only its cloud but also the hub
and user control interfaces (e.g., web and mobile app).
Moreover, many IoT platforms share users' data with partners (e.g., advertisers) 
for the expansion of businesses \cite{wink2018privacypolicy, smartthings2018privacypolicy, vera2018privacypolicy}; any improper protection may exfiltrate private data to third parties. 

Our investigation of popular smart home platforms shows that
these platforms are factually overprivileged to access real-time data streams from connected devices, 
although most of the data do not trigger any automation. 
This deviates the principle of ``data minimisation'' in European General Data 
Protection Regulation (GDPR) \cite{GDPR2016} or ``least privilege'' in access 
control systems \cite{fernandes2016security}. 
We also find that no capabilities are provided for  
users to control the leakage of private device data to the platform, failing to realize user-centric authorization. Therefore, our goals are to minimize the data sent to the platform and allow users to
define customizable data flow control policies for individual privacy preferences.

Multiple challenges arise for attaining these goals. 
First, the data minimization should not \emph{adversely} affect the functionality of home automation 
.
We observe that the semantics of home automation apps can be represented as rules with each following a \emph{event-condition-action} model and the state-of-art code analysis techniques~\cite{chi2018cross,celik2018soteria,tian2017smartauth} are proved to be effective in extracting rule semantics from apps. Our insight is that by finding the minimum data flows required by these rule semantics, we can properly generate and enforce data flow control policies without affecting home automation. For example, suppose a rule has a semantic ``\emph{when a motion is detected, if the indoor temperature is higher than 79$^{\circ}F$, turn on the A/C}''.
We can convert it into a data-minimization policy, such that if the indoor temperature is not higher than 79$^{\circ}F$,
no data is sent to the platform; besides, if the A/C is already on (that is, the rule execution does not
change anything), no data
is sent even if the temperature is higher than 79$^{\circ}F$. Optionally, users can have the system 
fuzz the data, such that even if the policy execution determines that the temperature 
should be sent, a random value larger than 79 is reported.

Second, many platforms are closed systems that do not allow platform-level modifications and
it is probably unrealistic to expect a platform to cooperate to enforce 
data minimization. Thus, how to enforce data-protection policies before data leave the home network is a challenge. Intuitively, one may propose to \emph{circumvent} this challenge
by building a new purely-local platform, such that no data have to flow out of a home; or, one can simply cut the network cable of a local gateway~\cite{mozillaiot2019} and enforce most of the home automation locally.
However, a large number of existing platforms have been deployed in homes and it might be infeasible to convince users to switch to another new platform they are not familiar with; moreover, a purely local platform means that a lot of highly desired Internet-based services (e.g.,
messaging, storage, and remote management) will be cut out. Therefore,
how to enforce data protection on the existing platform architecture
without sacrificing the values of Internet-based services imposes extra difficulties.

We leverage multiple system-building ideas into our system, named \tool. 
First, we build \tool as a data mediator,
which sits between IoT devices and the hub to transparently filter data based on
privacy-protection policies. The advantage is that neither IoT devices
nor the platform needs to be modified. Thus, another challenge is that
the original communication between IoT devices and the hub is encrypted,
which prevents \tool from understanding and then filtering data.
We overcome this difficulty with a man-in-the-middle approach:
the data mediator claims itself as a hub to pair
with all the devices, and meanwhile it creates the same number of virtual devices to connect the hub. 

Furthermore, we borrow the idea of a DMZ (demilitarized zone) when designing
\tool. A DMZ exposes certain external-facing services (e.g., web)
to the Internet, while the organization's local area network (LAN) is segregated 
by a firewall. This way, even a node in the DMZ is compromised, attackers need to 
bypass the firewall to reach the LAN. We propose to place the hub in a DMZ, and set 
up an extremely simple firewall between the DMZ and \tool: the external world cannot
initiate connection to \tool, and any inbound traffic, unless it targets those virtual
devices, should be discarded immediately by \tool. 

We demonstrate the ideas by implementing \tool to work with two representative platforms: Samsung SmartThings and openHAB, which
are of the most popular \emph{cloud-based} and \emph{gateway-based} IoT platforms, respectively. We evaluate \tool in two real-world deployments. The results suggest that \tool reduces the amount of data sent to the platform
by 97\% based on data-minimization policies. Our case study shows that the data 
reduction heavily impairs the attacker's 
ability to infer privacy-sensitive behaviors, e.g., bathroom usage 
and the arrival and departure time of home members. The user-specified
policies provides extra fine-grained data control to resolve personalized privacy preferences and concerns.

The contributions of this work are summarized as follows.
 
\begin{itemize}[leftmargin=*]
	\item We reveal the fact that most smart home platforms employ a simple trust-by-default model between home devices and the platforms, resulting in over-leakage of sensitive IoT device data. We find several channels through which the collected data could be revealed, demonstrating the severe privacy risks. Despite the clear need for user-centric data flow control, we find that most leading platforms do not have supports for this purpose.   

	\item We design an effective data flow control system to enhance user privacy in home automation. On one hand, data-minimization policies are automatically generated based on the installed automation apps, reporting minimally necessary data for app execution and obfuscating the reported data for further protection. On the other, users are offered capabilities to prioritize policies specified by themselves to customize data flow control for individual privacy preferences and concerns. 
	
	\item A man-in-the-middle style enforcement mechanism in closed-source smart home systems is designed. A proxy device mediates the communication between IoT devices and the hub, without modifying the devices or the hub. 

	\item We implement a proof-of-concept prototype to work with two platforms: SmartThings and openHAB. Through the evaluation in two real-world scenarios: a two-bedroom apartment and a public workplace, we demonstrate that our system significantly reduces the privacy risks due to data leakage and introduces negligible latency to home automation. A user study is conducted to learn users' attitude and capabilities towards defining privacy-protection policies with mobile interfaces. 
\end{itemize}


\section{Background: Smart Home Platforms}
\label{section_background}
\begin{figure}[t] 
	\centering
	\includegraphics[width=0.47\textwidth]{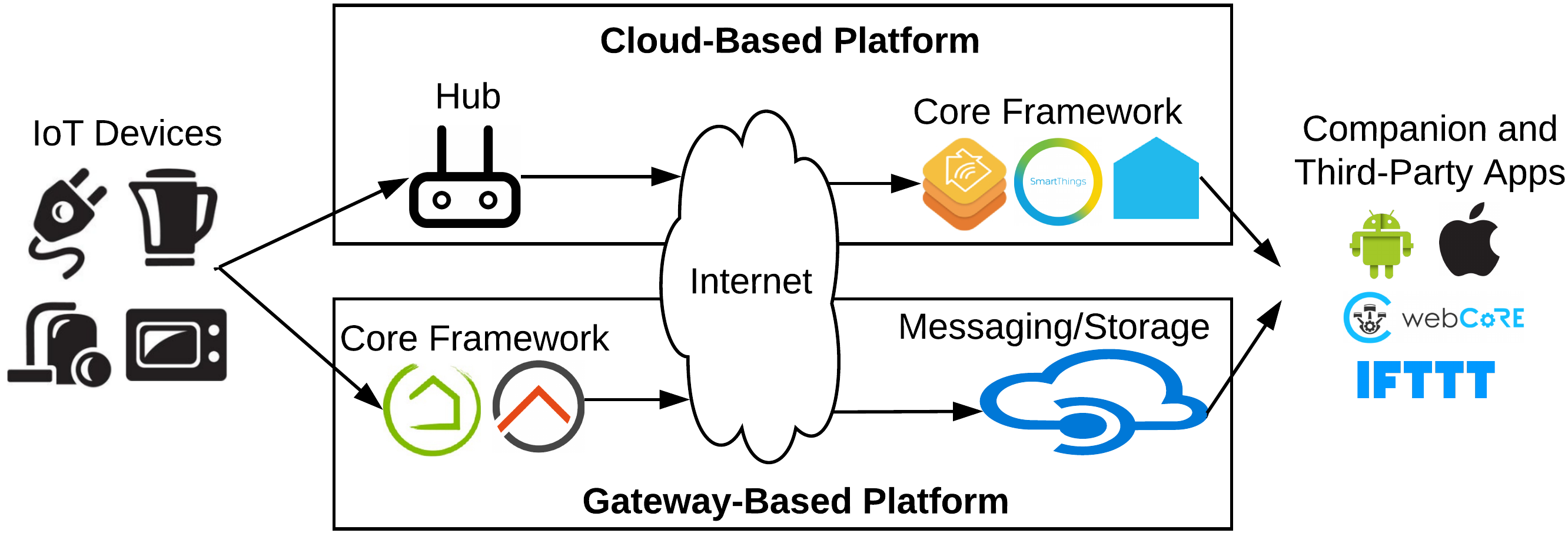}\\ 
	\caption{Smart home platform architecture.}\label{fig_smartthings} 
	\vspace{-10pt}
\end{figure}

Smart home platforms can be categorized into two types: cloud-based platforms (CBPs) and gateway/hub-based platforms (GBPs), according to whether the core framework of a platform is hosted in a remote cloud or a gateway/hub device located at home (as shown in Fig.~\ref{fig_smartthings}); the two types are similar, otherwise. 
Note that the gateway running a
core framework at home does not resolve the privacy leakage threats, as the
gateway connects to the Internet and is under the full control of the
platform administrator. Once the platform is compromised, the attacker
gains equivalent capabilities of gaining user data. 
We choose a CBP---\emph{SmartThings}, one of the most popular and full-fledged 
platforms, as an example to describe the key components in a smart home system.

\begin{itemize}[leftmargin=*]
\item \textbf{Hub.}
A CBP hub connects IoT devices through distinct short/medium-range wireless radios (ZigBee, Z-Wave, etc.). 
The hub plays a key role to ensure the interconnectivity and interoperability of heterogeneous IoT devices. A GBP also has a hub-like device\footnote{We use \emph{gateway} for distinguishment.} which not only connects IoT devices but also hosts the core framework (descrbied below). Note that the hub or gateway device, though physically located at home, is conceptually regarded as a part of the platform in terms of data privacy protection in that it is under the fully control of the platform administrator.

\item \textbf{Cloud.}
The backend cloud of a CBP hosts the core framework and provides cloud messaging, storage as well as any other necessary services for the platform to function. The cloud in a GBP is typically responsible for messaging and storage. The cloud messaging service facilitates some critical functionalities, such as notification, third-party application integration, remote monitoring and control.
Many Internet-based services depend on the cloud. 

\item \textbf{Core Framework.}
The core framework runs major functionalities of a platform, including home automation. Take SmartThings as an example. It provides a sandboxed runtime environment for running \emph{device handlers} and \emph{SmartApps}. Device handlers are software wrappers of physical devices which abstract the physical devices
(as a set of \emph{capabilities} and handle the underlying protocol-specific communications between the core framework and the physical devices).
They expose uniform interfaces for SmartApps to interact with devices. 

\item \textbf{Companion and Third-Party Apps.}
To provide a convenient user interface (UI) for users to manage their hubs, IoT devices 
and apps, a platform usually provides a smartphone \emph{companion} app. 
For instance, in SmartThings companion app, users can install and configure a SmartApp.
Current platforms also expose interfaces (mostly RESTful cloud APIs) to incorporate third-party services/applications (e.g., mobile apps, IFTTT \cite{ifttt}, webCoRE \cite{webcore}).
\end{itemize}

Therefore, a smart home platform has a large attack surface involving
the hub, cloud, core framework services, companion app, and APIs for third
parties, let alone inside attacks. It is dangerous and unnecessary that users grant unlimited trust to it by allowing all the data to flow to the platform. 

\section{Motivation and Threat Model}
In this section, we first reveal two facts we have observed, and then present the threat model.

\subsection{Privacy Concerns about Platforms}
\label{subsec_privacy_implications}
\subsubsection{Trust By Default}
In smart home systems, the platforms are typically fully trusted. That said, after being installed, a platform gains the access privilege to all connected home devices technically by design and legally by claiming a \emph{terms and conditions} or a \emph{privacy policy}. To reduce development complexity and the time to market, most emerging platforms do not provide access control between home devices and their hubs to avoid accessing unnecessary data; instead, they simply collect all data streams reported by devices for further processing. We studied the privacy-related practices in popular smart home platforms and showed the details in Appendix~\ref{section_investigation}. In this section we use SmartThings as an exemplar to demonstrate.

\vspace{3pt}
\textbf{Are home data flowing out of homes silently?} 
To answer this question, we connected four types of ZigBee devices (a multipurpose sensor, a motion sensor, an arrival sensor and an outlet) and a Z-Wave sensor (Aeotec Multisensor 6) to a SmartThings hub and inserted \texttt{log.debug} code into the \texttt{parse} methods of \texttt{device handlers} which are used by the core framework to parse the received IoT payload and generate in-system events. In this way, we obtain all data received by the SmartThings cloud via its hub on the living logging interface \cite{smartthings2018ide}. We did \emph{not} install any automation apps and did \emph{not} operate any SmartThings-provided interfaces; we only interacted with the devices physically. We found that the platform cloud still kept receiving device attribute data (e.g., motion, switch, temperature, etc.) from the above devices, indicating that device data flow out via the hub even if they are not subscribed to or requested by any service.


This trust-by-default model introduces severe data leakage risks to smart homes since attackers may gain unauthorized access to home data by compromising the hub device, cloud infrastructure, or the companion app~\cite{alrawi2019sok}. Vulnerabilities in IoT platforms and clouds have been demonstrated by recent works. For instance, Fernandes et al. \cite{fernandes2016security} and Zuo et al. \cite{zuo2019does} respectively revealed that the abuse of OAuth tokens and cloud API tokens in mobile apps imposes significant security and privacy threats including unauthorized access to the platform. An inside attacker can also access all the data.


\subsubsection{Limited User Capabilities}
Users visibility and control helps mitigate risks. However, users have few capabilities and interfaces to inspect or control what their device sends to the Internet~\cite{alrawi2019sok}. They only have a binary choice: whether or not to connect a device to the platform; once connected, the device keeps reporting data to the hub device continuously and opaquely.

\subsection{Threat Model}
We consider the platform may be
exploited by attackers for accessing user private data and inferring user privacy-sensitive behaviors. Attacks that exploit the home IoT device hardware vulnerabilities, side channels, or home local networks to steal private data are out of the scope of this work. We assume the home automation apps are not 
malicious (note that how to detect and handle malicious automation apps 
is a separate problem and has been well studied, e.g.~\cite{tian2017smartauth, celik2018soteria, ifttt}).

\begin{figure}[tb]
	\centering
	\includegraphics[width=0.47\textwidth]{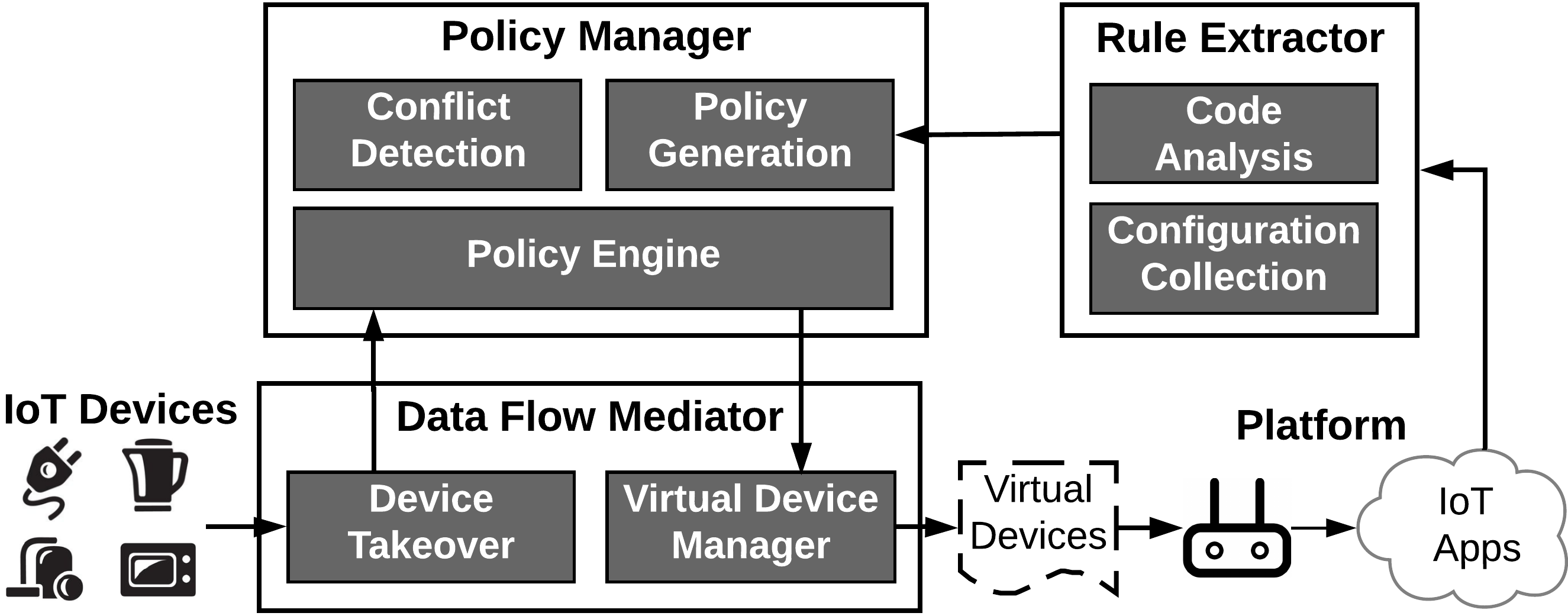}\\
	\caption{The architecture of \tool.}\label{fig_architecture}
\end{figure}

\section{\tool System Overview}
To mitigate data leakage, we propose to introduce access control before data leave the control of users. In this way, privacy-oriented metrics can be applied to provide the data exposure with certain privacy guarantees (i.e., data minimization in this paper) and end-user controls are also feasible to satisfy personal privacy preferences. However, it is challenging to attain these goals for the following reasons. 
First, the data filtering, if not carefully performed, may accidentally affect
home automation. Thus, how to precisely analyze apps and convert them
into privacy-protection policies correctly is a challenge (Section~\ref{section_policy_system}).
Most smart home platforms
are closed-systems and do not allow platform-level modifications. Moreover,
the traffic between IoT devices and the hub is encrypted. How to perform
the data filtering without modifying the device, hub, or platform framework is challenging 
(Section~\ref{section_relay}). How to provide interfaces for non-expert 
users to define their own privacy-protection policies is non-trivial (Section~\ref{section_policy_generation}).

For interoperability, the wireless protocols in IoT devices are mostly open-source standard ones such as ZigBee, Z-Wave, LAN, etc., which makes it possible to place a man-in-the-middle device (named \emph{mediator}) between IoT devices and the hub to intervene in the communication between them. On top of the mediator, it becomes possible to process the raw data flows before forwarding them to the hub. With this insight, we build \tool, a system that enforces carefully generated data flow control policies before data are reported to the backend platform for home automation.
As shown in Fig.~\ref{fig_architecture}, \tool comprises the following modules:

\begin{itemize}[leftmargin=*]
	\item \textbf{Rule extractor} extracts the home automation rules from rule creation interfaces, e.g., IoT apps, webpages, smartphone apps, etc. When rules are initially installed, the rule extractor obtains rule semantics and rule-device binding information. In appified IoT systems, the rule extractor comprises a \emph{code analysis} component to extract rule semantics from apps and a \emph{configuration collection} component to collect rule-device binding information. The rule semantics and rule-device bindings constitute the complete automation logic.
	
	\item \textbf{Policy Manager} generates and manages data flow policies used for protecting IoT data. \emph{Policy generation}, on one hand, interacts with the rule extractor to generate semantics-based data-minimization policies; on the other hand, it takes in user-specified policies from the user interfaces and formats them into executable-formatted policies. \emph{Conflict detection} inspects if a user-specified policy conflicts with existing data-minimization policies and thus affects home automation; when conflicts are detected, it reports the conflict to the user for making decisions. \emph{Policy engine} interprets and executes the above policies over the incoming raw data from IoT devices. 
	
	\item \textbf{Data Flow Mediator} is a proxy who mediates the communication between IoT devices and the hub. The mediator, on behalf of the hub, talks with IoT devices via device-dependent protocols (e.g., ZigBee, ZWave, WiFi, etc) and forwards the raw device data to the \emph{policy engine} for processing. On the other hand, the mediator creates a virtual device instance to send the processed data to the hub, on behalf of each real device. All virtual device instances use a uniform communication protocol supported by the target platform (e.g., LAN in SmartThings \cite{lanconnected2018} and MQTT in openHAB \cite{mqtt2019}). Besides, the virtual devices receive device control commands from the hub, which will then be translated to protocol-specific commands and forwarded to the corresponding real device. The data mediation is not transparent to the platform and therefore the platform works exactly the same way.
\end{itemize}


\section{Design and Implementation}
\label{section_design_implementation}
In this section, we present the detailed design and implementation of \tool. 
We choose Samsung's SmartThings, one of the most mature and comprehensive smart home platforms, as the underlying platform to describe the implementation of \tool. We first describe our policy generation and management for contextually controlling IoT data flows. Then, we present how we enforce policies in existing IoT systems by introducing a data flow mediator. To show the applicability of \tool, we also present how we integrate \tool with another platform, openHAB, by adapting the platform-specific components.

\subsection{Data Flow Control Policies}
\label{section_policy_system}
\subsubsection{Policy Definition and Execution}
\label{section_policy_definition}
Home automation is context-aware: a rule executes a command when it is triggered by an \emph{event} and meanwhile the smart home is under the prescribed \emph{condition}. Note that the event and condition are slightly different: an event describes a context change (e.g., the motion sensor's reading changes from ``inactive'' to ``active'', which indicating a motion is detected) while a condition indicates a collection of static statuses (e.g., the motion sensor's latest reading is ``active''). To precisely filter raw IoT data flows for data minimization without interfering with the execution of automation rules, data flows need to be processed contextually. 
To this end, we define a context-aware policy format.

\begin{lstlisting}[caption={Context-aware policy format},label=listing_policy,captionpos=b, abovecaptionskip=2pt, numbers=right, float,floatplacement=H]
TRIGGER:{
	match (:type).(:subject).(:attribute)
	satisfy (:operator)->(:value)
	[fetch1] (:type).(:subject).(:attribute*)
	[branch] (:operator1)->(:value)
	run (:method)(:parameters)(:delay)
	[else] (:method1)(:parameters1)(:delay1)
}
CHECK: [{
    fetch (:type).(:subject).(:attribute)
	satisfy (:operator)->(:value)
	[fetch1] (:type).(:subject).(:attribute*)
	[branch] (:operator)->(:value)
	run (:method)(:parameters)
	[else] (:method1)(:parameters1)
	}, ...]
\end{lstlisting}

Formally, we define a data flow policy as $\mathbf P\tt \mathbf=(T, C)$, where $\tt T$ and $\tt C$ denote the \texttt{TRIGGER} and \texttt{CHECK} section in a policy as shown in Listing \ref{listing_policy}. \texttt{TRIGGER} defines the incoming event that triggers the execution of $\mathbf P$ and \texttt{CHECK} encapsulates a list of items, each of which indicates a constraint that must be satisfied for the policy to indeed perform actions. \texttt{type} indicates that the event is fired by a device or is a time change, etc; \texttt{subject} is to identify a specific IoT device (i.e., device ID); {attribute} specifies the attribute of a device (which may have multiple attributes) or the time-related feature (e.g., time of day, date, timer). \texttt{type}, \texttt{subject} and \texttt{attribute} are to check if an incoming data matches the event that triggers the policy in \texttt{TRIGGER} and are to query the smart home status for constraint checking in \texttt{CHECK}. \texttt{operator} and \texttt{value} denote a constraint that the incoming event or smart home status must satisfy for the policy context to be evaluated as true. A policy action defined in the \texttt{run} fields where \texttt{method} and \texttt{parameters} define how to process the raw data and \texttt{delay} controls the timing for reporting the processed data to the platform. Besides, there are three optional fields marked with ``[]'' that form an extended \texttt{TRIGGER} section or a \texttt{CHECK} item. [\texttt{fetch1}] and [\texttt{branch}] evaluate an extra constraint on the fetched data; if true action defined in \texttt{run} is executed, and otherwise action in \texttt{else} will be executed instead.

Policies are executed by a policy engine. The policy engine listens to all the incoming raw data from the IoT devices and time-related information if registered. When receiving a new data item $\texttt D$ (a.k.a. an event), the engine uses $\texttt D$ to evaluate the maintained data flow policies one by one. Algorithm \ref{alg_policy_engine} shows the general workflow of how the engine evaluates and executes a policy $\mathbf{P}$. Specifically, it first checks if $\tt D$ matches the \texttt{type}, \texttt{subject}, and \texttt{attribute} in \texttt{TRIGGER}, and then examines if the value of $\texttt D$ satisfies the constraint specified by \texttt{operator} and \texttt{value}. If true, $\mathbf P$ is triggered and proceeds to execute. Then the engine evaluates all items specified in \texttt{CHECK}. Since the data required for evaluating the \texttt{CHECK} items are not newly captured events but the current smart home status (e.g., the device working status), the policy engine fetches the information indexed by \texttt{type}, \texttt{subject} and \texttt{attribute} from a database \texttt{DB}, which stores the latest attribute values of all connected devices and updates them when devices report any change. Only when constraints defined in all \texttt{CHECK} items are satisfied, the policy is finally evaluated and the actions defined in all \texttt{run} or \texttt{else} fields will be performed. During the above process, a policy terminates if there is any event mismatches or constraint violation. Besides, the policy engine also maintains another database \texttt{DB}$^{*}$ to keep record of the lastest reported data for each device attribute.

\begin{algorithm}
	\scriptsize
	\SetKwInOut{Inputs}{Input}\SetKwInOut{Output}{Output}
	\Inputs{$\tt D \leftarrow$ new data item, $\tt P \leftarrow$ A privacy policy \\ $\tt DB\leftarrow$ Newest Device Status Database \\ $\tt DB^{*}\leftarrow$ Newest Reported Data Database}
	\Output{Privacy-Aware Data Set $\tt DS$}
	
	\If{match($\tt D.source$, $\tt P.TRIGGER.(type, subject, attribute)$) and satisfy($\tt D.value$, $\tt P.TRIGGER.(operator, value)$)} {
		\ForEach{ {\tt checkitem}$\in${\tt P.CHECK} } {
			{\tt val} $\leftarrow$ {\it fetch} ($\tt DB$, $\tt checkitem.(\tt type, subject, attribute)$) \\
			\If{!satisfy($\tt val$, $\tt checkitem.(\tt operator, value)$)} {
				return\\
			}
		}
		\eIf{ {\tt P.TRIGGER}.contains({\tt [branch]}) } {
		    $\tt val^{*} \leftarrow${\it fetch} ($\tt DB^{*}$, ${\tt P.TRIGGER}.(\tt type,subject,attribute^{*})$) \\
	    	\eIf{satisfy($\tt val^{*}$, \tt P.TRIGGER.(operator1,value))} {
			    $\tt DS \leftarrow$ {\bf run} $\tt P.TRIGGER.(method, paras., delay)$\\
		    } 
            {
		        $\tt DS \leftarrow$ {\bf run} $\tt P.TRIGGER.(method1, paras.1, delay1)$\\
		    }
		} 
        {
		    $\tt DS \leftarrow$ {\bf run} $\tt P.TRIGGER.(method, parameters, delay)$\\
		}
		
		\ForEach{$\tt checkitem\in \tt P.check$} {
			\eIf{ {\tt checkitem}.contains(\tt [branch])} {
				$\tt val^{*} \leftarrow${\it fetch} ($\tt DB^{*}$, $\tt checkitem.(\tt type,subject,attribute^{*})$) \\
				\eIf{satisfy($\tt val^{*}$, $\tt checkitem.(\tt operator, value)$)} {
				    $\tt DS \leftarrow$ $\tt checkitem.(method, paras.)$\\
			    } {
			        $\tt DS \leftarrow$ $\tt checkitem.(method1, paras.1)$\\
			    }
			} {
			    $\tt DS \leftarrow$ $\tt checkitem.(method, paras.)$ \\
			}
		}
	} 
	\caption{The algorithm for executing a policy}
	\label{alg_policy_engine}
\end{algorithm} 

\subsubsection{Policy Generation}
\label{section_policy_generation}
\tool generates two types of policies: automation-based data-minimization policies (APs) and user-specified policies (UPs). To achieve data minimization, i.e., only report the minimum amount of data that are necessary for home automation, rules are extracted from installed automation apps and analyzed to find the minimum data flows for the rules to execute. UPs are generated from user interfaces and work with APs simultaneously, which is an important supplement to customize privacy preferences that cannot be learned from home automation.

\vspace{10pt}
\noindent\textbf{Automation Rule Extraction}\vspace{3pt} \\
Rule extraction is the first step for AP generation. Automation rules follow an event-condition-action model and are installed by installing IoT apps or selecting rule templates on web or mobile app interfaces. The rule extraction regarding both methods has been widely studied by state-of-art literature. Code analysis has been proved to be an effective way to extract rule semantics from IoT apps by state-of-art work. For example, by utilizing Abstract Syntax Tree (AST) analysis on smart apps, \cite{fernandes2017internet} identifies requested and used capabilities in SmartApps, \cite{tian2017smartauth, ding2018safety} breaks down SmartApps and extracts rule information, \cite{jia2017contexiot, zhang2018homonit, celikiotguard} builds Deterministic Finite Automatons (DFAs) from SmartApps. Symbolic execution is a more powerful technique to analyze rule semantics from apps \cite{chi2018cross, celik2018soteria}. Text data crawling and natural language processing (NLP) are used for rule extraction from web pages and mobile apps \cite{zhang2018homonit, hwang2016data}.  

\begin{figure}[tb]
	\centering
	\includegraphics[width=0.46\textwidth]{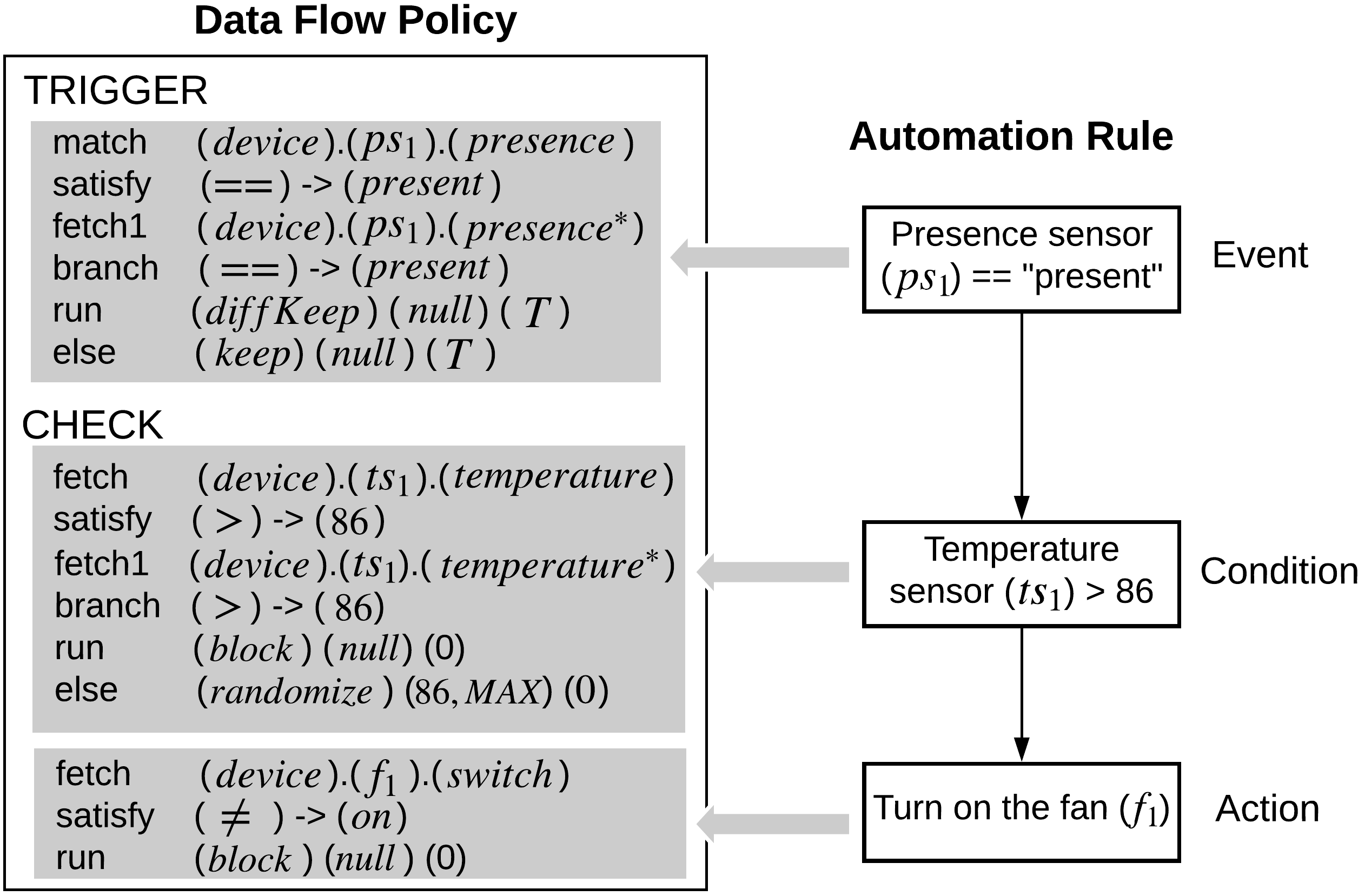}\\
	\caption{The policy derivation from an automation rule.}\label{fig_policy_from_rule}
	\vspace{-10pt}
\end{figure}

Rather than design another code analyzer, in this paper, we adapt the solution provided in \cite{chi2018cross} to implement our rule extractor since it not only implements a complete symbolic executor with API modeling but also provides an app-device binding collection approach. We obtain the source code from the authors and verify its effectiveness on 86 SmartApps from SmartThings market apps. The executor works on the AST representation of a SmartApp; the rule extraction starts from an event subscription method \texttt{subscribe()} (event that triggers a rule) and traces in the entry point of the event handler method. All paths branching at \texttt{if-else} statements (rule condition) are explored until a \texttt{sink} (rule action) is spotted; expressions (e.g., value assignment) and APIs (e.g., device access methods, device control commmands) along the paths are modelled\footnote{Due to page limits, we refer interested readers to the literature \cite{chi2018cross} for more details.}. The combination of control flow analysis and data flow analysis allow us to extract the rule context (event and condition) and command (action) from a SmartApp. The right column of Fig.~\ref{fig_policy_from_rule} shows the extracted rule from a temperature control SmartApp that defines a rule $\mathbf{R}_{1}$ ``when a presence sensor $ps_{1}$ becomes \emph{present}
, if the reading of a temperature sensor $ts_{1}$ 
is higher than 86$^{\circ }F$, turn on the fan $f_{1}$''.

\vspace{3pt}
\noindent\textbf{Data-Minimization Policy Generation}\vspace{3pt} \\
Consider the example rule $\mathbf{R}_{1}$. By default, the platform continuously receives and stores data streams from devices (presence sensor, temperature sensor, fan).  
However, we observe that these data are not all required for executing $\mathbf{R}_{1}$ in cases:

\begin{enumerate}
\renewcommand\labelenumi{(\theenumi)}
    \item The presence sensor $ps_{1}$ does not send any event;
    \item $ps_{1}$ sends a ``not present'' event; 
    \item The indoor temperature measured by $ts_{1}$ is lower than 86$^{\circ }F$;
    \item The fan $f_{1}$ is ``ON'';
    \item $ps_{1}$ sends a ``present'' event and the last reported temperature by $ts_{1}$ is higher than 86$^{\circ }F$. 
\end{enumerate}

In cases (1)-(4), there is no need to report any data from $ps_{1}$ and $ts_{1}$ to the platform; in case (5), it is unnecessary to report temperature data since the temperature value stored in the platform database satisfies the rule condition checking; in no cases, the ON/OFF state of $f_{1}$ is useful for executing $\mathbf{R}_{1}$. From this example, we can conclude that only sporadic ones in the data streams of devices are required for home automation, which motivates us to encode highly-structured automation rules to data-minimization policies. An example of generating an AP from $\mathbf{R}_{1}$ is shown in Figure~\ref{fig_policy_from_rule}. The \texttt{TRIGGER} of AP is derived from the \texttt{Event} of $\mathbf{R}_{1}$ and \texttt{CHECK} is derived from the \texttt{Condition} and \texttt{Action} of $\mathbf{R}_{1}$, respectively. According to the policy definition and execution algorithm presented in Section~\ref{section_policy_definition}, the derived AP expresses multi-faceted information for \tool to process data:

\begin{enumerate}[leftmargin=*]
    \item Context: when and only when an incomming event of $ps_{1}$ is ``present'' and meanwhile the latest received reading of $ts_{1}$ is higher than 86$^{\circ }F$ and the state of $f_{1}$ is not ``ON'', \emph{some} data will be reported, and otherwise, the policy will be skipped and no data will be reported at all;
    \item Event reporting: if the latest reported value of $ps_{1}$ is ``present'', use the \texttt{diffKeep()} method to process the current value for reporting, and otherwise, use \texttt{keep()}; 
    \item \texttt{CHECK} data reporting: if the latest reported value of $ts_{1}$ is higher than 86$^{\circ }F$, use the \texttt{block()} method to process the current value of $ts_{1}$, and otherwise, use \texttt{randomize(86, MAX)}; use \texttt{block()} to process the state data of $f_{1}$.
\end{enumerate}

Table~\ref{table_methods} shows a summary of all the methods used in the \texttt{run} and \texttt{else} fields. In the default setting, binary sensors such as the presence sensor reports binary values alternatively; thus, SmartThings only fires an event when observing a value change. Our data flow control breaks the alternate ``present'' and ``not present'' values in the data stream of $ps_{1}$. 
Thus, when the platform receives ``present'' but finds the last value is also ``present'', it will not issue a ``present'' event in its framework and $\mathbf{R}_{1}$ cannot be triggered. Hence, the derived AP uses \texttt{diffKeep()} rather than \texttt{keep()} to address this issue; \texttt{diffKeep()} reports ``not present'' followed by ``present'' with a time delay $T$, which ensures a ``present'' event is fired. It is worth mentioning that the selection of $T$ is non-trivial to guarantee the normal execution of home automation because it allows time for the platform to update a received data to its database. Similarly, it is required that SmartThings have updated the temperature value (if necessary) in database before it issues a ``present'' event to $\mathbf{R}_{1}$; otherwise, the app will fail the temperature condition check when triggered by the event\footnote{We manually observed app execution while tuning $T$ and found a value as small as 100 millisecond without causing failure in 1000 trials.}. The \texttt{block()} discards data without sending it. \texttt{randomize()} randomizes the float-value attribute data (e.g., temperature). In the example, the temperature is used to compare with a threshold (86$^{\circ }F$), so a random value between 86$^{\circ }F$ and the upper limit of a temperature $MAX$ is sufficient for the condition checking. \texttt{MAX}/\texttt{MIN} denotes the upper and lower boundaries of a specific attribute (See Table~\ref{table_boundary_values}). We obtain such information from SmartThings Capabilities Reference \cite{smartthings2018capability}. Besides, we present how \tool handles time/timer-related automation in Appendix~\ref{appendix_time_timer}. 

\begin{table}
	\scriptsize
	\caption{Summary of methods used in data flow policies}
	\renewcommand\arraystretch{1}
	\newcommand{\tabincell}[2]{\begin{tabular}{@{}#1@{}}#2\end{tabular}}
	\newcolumntype{P}[1]{>{\arraybackslash}p{#1}}
	\newcolumntype{M}[1]{>{\centering\arraybackslash}m{#1}}
	\begin{tabular}{P{2.5cm}P{5cm}}		
		\toprule
		\bf Method & \bf Description\\ \midrule
		\texttt{keep()} & Report the original value\\ 
		\texttt{block()} & Do not report\\
		\texttt{diffKeep()} & Report a different value and then the original value\\ 
		\texttt{randomize(MIN,MAX)} & Report a random value $\in$ (\texttt{MIN},\texttt{MAX})\\ 
		\texttt{pickOther(CUR,ENUM)} & Randomly picked a value ($\neq$\texttt{CUR}) from set \texttt{ENUM} \\ 
		\bottomrule 
	\end{tabular}
	\label{table_methods}
\end{table}

\begin{table}
	\centering
	\scriptsize
	\caption{Boundary values for randomizing different attributes}
	\renewcommand\arraystretch{1}
	\newcommand{\tabincell}[2]{\begin{tabular}{@{}#1@{}}#2\end{tabular}}
	\newcolumntype{P}[1]{>{\centering\arraybackslash}p{#1}}
	\newcolumntype{M}[1]{>{\centering\arraybackslash}m{#1}}
	\begin{tabular}{M{2cm}P{1.4cm}P{1.8cm}M{1.4cm}}		
		\toprule
		\bf Attribute & \bf Min & \bf Max & \bf Unit\\ \midrule
		Temperature & -50 & 150 & $^{\circ}$F\\ 
		Illuminance	& 0 & 100000 & Lux\\ 
		Humidity	& 0 & 100 & \% \\ 
		Power	& 0 & 1800 & Watt\\ 
	\bottomrule 
	\end{tabular}
	\label{table_boundary_values}
\end{table}

\begin{figure}[tb]
	\centering
 \subfigure[]{
    \label{fig_info} 
    \includegraphics[width=0.23\textwidth]{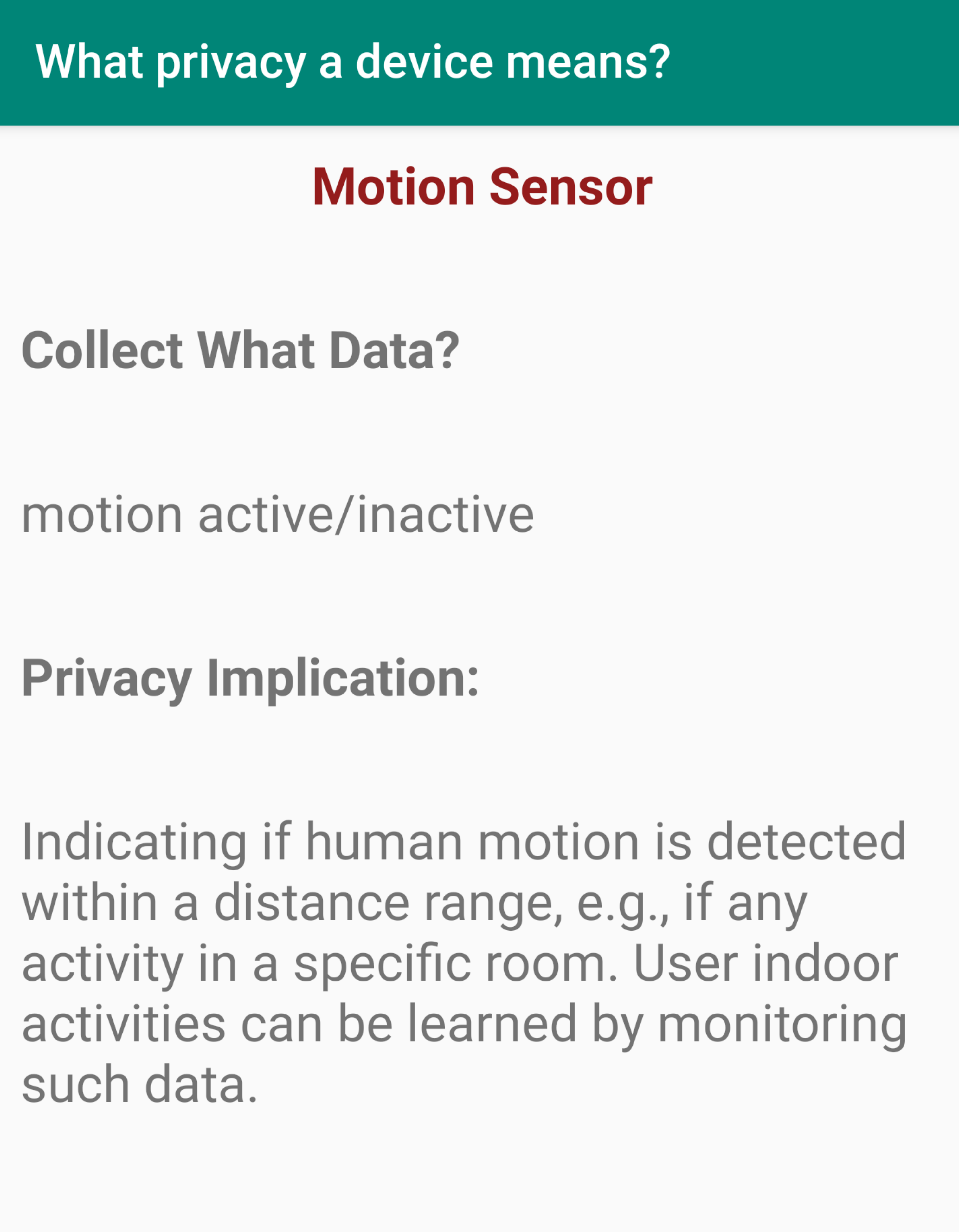}}
 \subfigure[]{
    \label{fig_policy} 
    \includegraphics[width=0.229\textwidth]{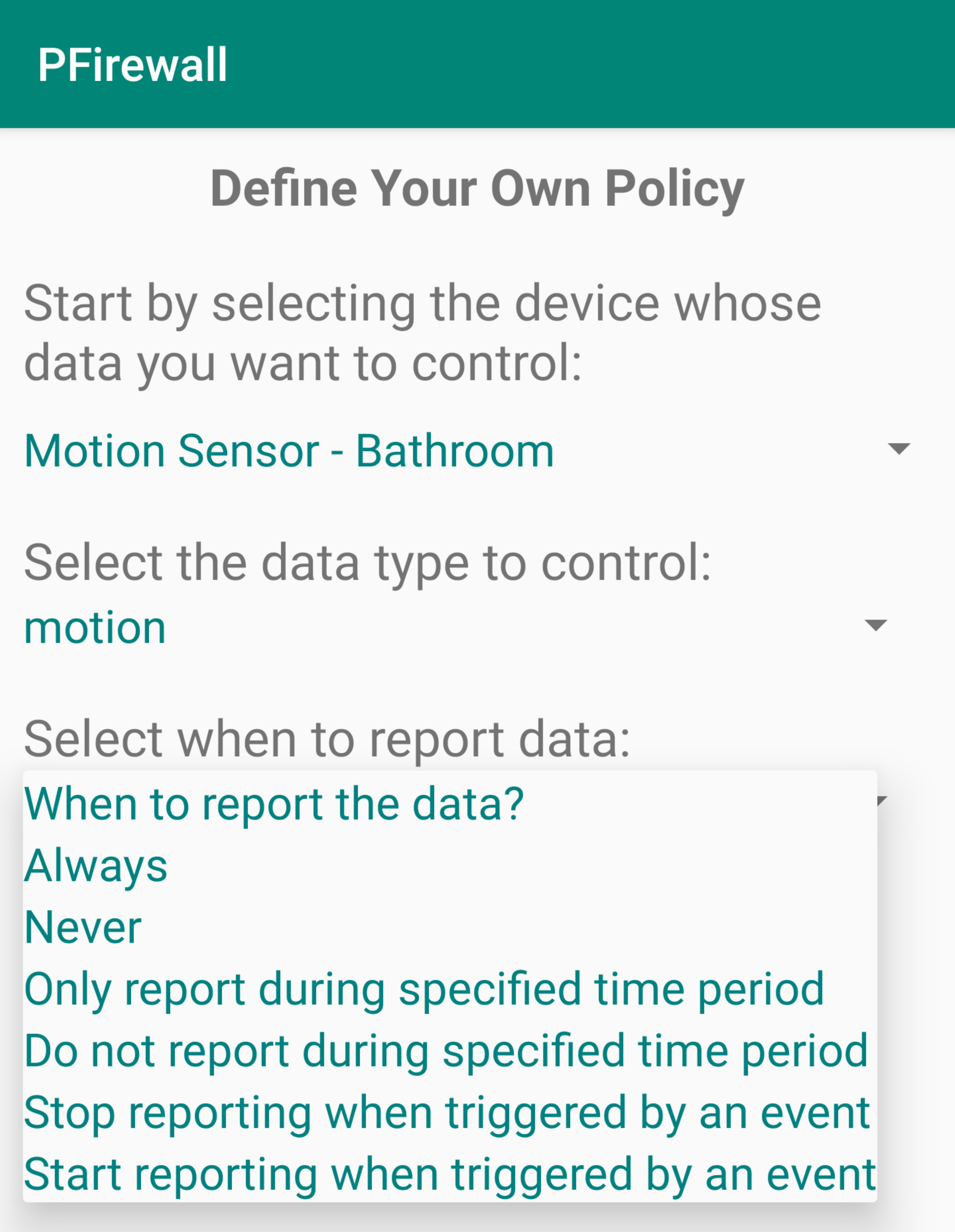}}
  \caption{Screenshots of \tool mobile app. The app provides an information tab showing users what data every device type generates and the corresponding privacy implications, and a policy tab allows users to define context-aware data control policies.}
  \label{fig_screenshot_app} 
	\vspace{-3pt}
\end{figure}

\vspace{3pt}
\noindent\textbf{User-Specified Policy Generation}\vspace{3pt} \\
We propose an interactive approach for users to specify data flow control policies. This is motivated by three reasons: 1) users have individual privacy preferences that cannot be derived from automation rules; for example, users might prioritize privacy rather than automation functionality for some device types during a time period or under certain situations; 2) the platform may integrate a third-party service but there is no rule extractor available to extract semantics from it; 3) users have rights to control the use of their data. In principle, UPs have higher priority than APs in controlling data. 

We develop a mobile app for end-users to specify policies. As shown in Fig.~\ref{fig_info}, information is displayed to help users understand what privacy issues each device and its data may imply. With the templates in Fig.~\ref{fig_policy}, users are able to configure whitelist, blacklist and conditional control policies during a specified time period or under certain contexts. Finally, UPs are encoded into the policy format in Listing~\ref{listing_policy} for execution. See Appendix~\ref{appendix_user_study} for the user survey we conducted to evaluate the policy templates. 

\subsubsection{Policy Conflicts}
A user is likely to define UPs which conflict with existing APs and hinder the automation since UPs are designed for overriding APs. Nevertheless, users need a warning that shows them what conflicts are imposed and which automation rules are affected. Therefore, an automated policy conflict detection is necessary. Two policies $\mathbf{P}_{1}$ and $\mathbf{P}_{2}$ \emph{conflict} if the following requirements are satisfied: 
(1) $\mathbf{P}_{1}$ and $\mathbf{P}_{2}$ are triggered simultaneously; i.e., an event makes both constraints $c_{T}^{1}$ and $c_{T}^{2}$ (defined in \texttt{TRIGGER} fields of $\mathbf{P}_{1}$ and $\mathbf{P}_{2}$, respectively) hold;
(2) both policies are finally executed i.e., all the constraints $c_{i}^{1}$ and $c_{i}^{2}$ in the \texttt{CHECK} fields of both policies are evaluated true;
(3) two policies define different actions (i.e., data processing methods, parameters, or delays) for the same data. Formally, let $\mathbf{S}(C)$ denote the set of all possible contexts that satisfy the set of constraints $C$, and $\mathbf{O}(a)$, $\mathbf{E}(a)$ denote the object (i.e., the controlled data) and effects of a certain action $a$ (defined in both \texttt{TRIGGER} and \texttt{CHECK} fields). A conflict occurs when the formula holds.

\begin{eqnarray}\label{equation_conflict}
\footnotesize
\left\{\begin{aligned}
&\mathbf{S}(c_{T}^{1}) \cap \mathbf{S}(c_{T}^{2}) \neq \emptyset, \\
&\mathbf{S}(c_{1}^{1}, c_{2}^{1}, \cdots) \cap \mathbf{S}(c_{1}^{2}, c_{2}^{2}, \cdots) \neq \emptyset, \\
&\exists i, j, \mathbf{O}(a_{i}^{1}) = \mathbf{O}(a_{j}^{2}), \mathbf{E}(a_{i}^{1}) \neq \mathbf{E}(a_{j}^{2}).
\end{aligned}
\right.
\end{eqnarray}

We detect policy conflict for each newly submitted UP against all APs. To calculate the constraint overlapping in the first two formulas in Equation~\ref{equation_conflict}, we encode each constraint in a policy into a quantifier-free first-order formulas: 
\vspace{-1mm}
\begin{align*}
\footnotesize
\underbrace{(\texttt{type}[.\texttt{subject}[.\texttt{attribute}]])}_\text{data source and type} (\texttt{operator}) (\texttt{value}).
\end{align*}
Thus, the constraint overlapping is transformed into a constraint satisfaction problem which can be solved by a constraint programming (CP) solver. In our implementation, we use a JavaScript linear solver \texttt{javascript-lp-solver} \cite{lpsolver}.

If the constraint satisfaction is solvable, two policies will be executed simultaneously. We then check whether the two policies perform different actions (by looking at the methods and parameters in \texttt{run} and \texttt{else} fields) on the same data flow; if so, the new UP conflict with an existing AP. The automation app which the AP was derived from would be affected and is displayed to users for making decisions.   

\subsection{Data Flow Mediation}
\label{section_relay}
\begin{figure}[tb]
	\centering
	\includegraphics[width=0.48\textwidth]{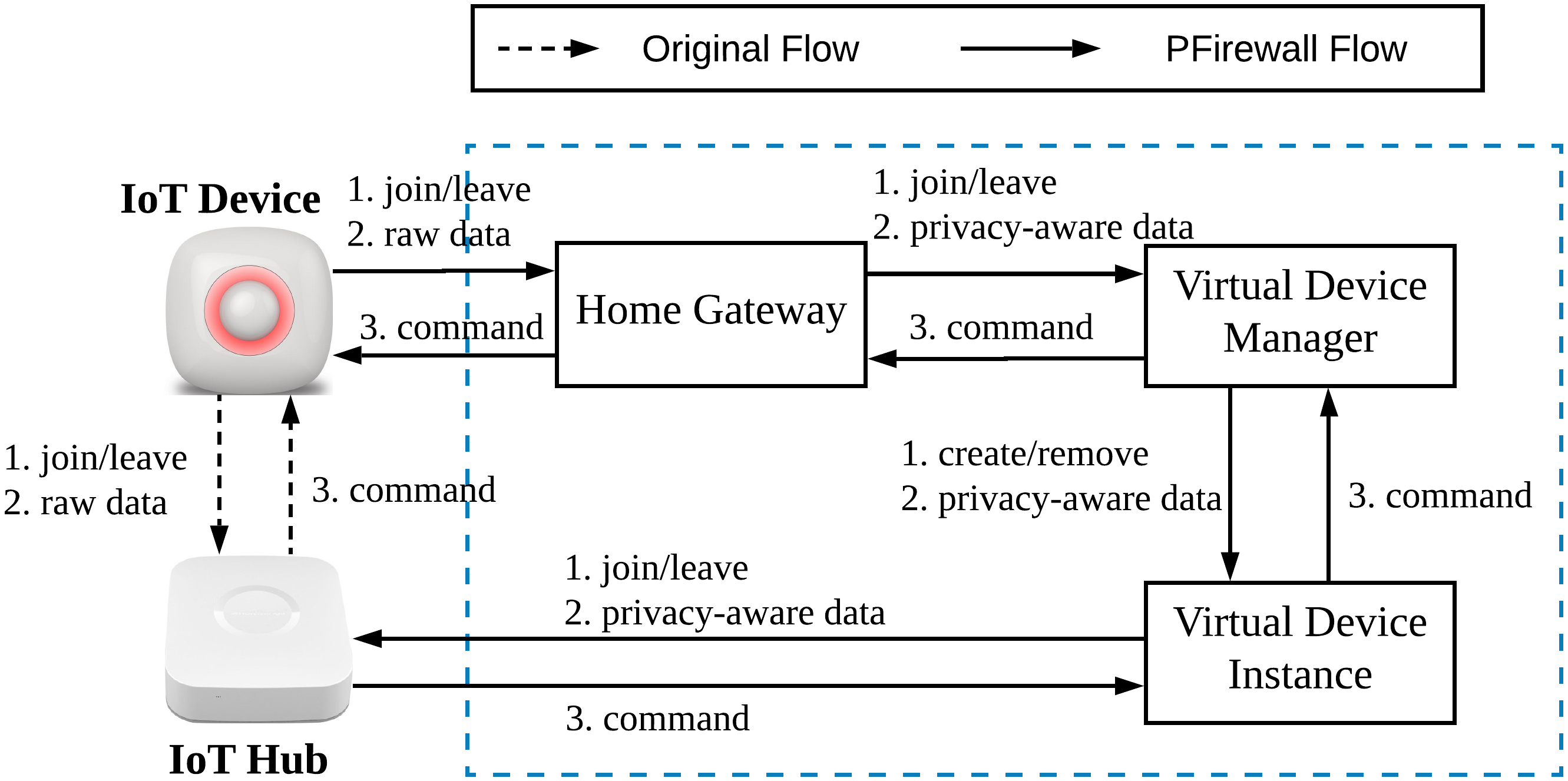}\\
	\caption{The workflow of the data flow mediator.}\label{fig_gateway_virtual}
\end{figure}

To enforce data flow policies in a closed-source IoT system, we introduce a \emph{data flow mediator} for relaying the communication between IoT devices and the hub, as shown in Fig. \ref{fig_gateway_virtual}. To this end, the mediator needs to (1) act as a hub to interact with IoT devices and (2) generate a virtual device to interact with the original hub on behalf of each real device.

\subsubsection{Connecting IoT Devices}
To play the role of a hub, the mediator needs to handle 3 major interactions with IoT devices: 1) devices join or leave the hub-leading network; 2) devices report attribute data to the hub; 3) the hub forwards commands from the platform to devices. The hub functionality is provided by many open-source platforms, e.g., openHAB \cite{openhab2019} and Mozilla IoT \cite{mozillaiot2019}, which allow developers to add add-ons for integrating various IoT devices using different communication techniques. Until now, openHAB supports 275 bindings that have been tested to work with hundreds of commercial IoT devices and Mozilla IoT also have tested more than 100 mainstream devices. In our implementation, we adapt the source code of Mozilla IoT to realize connecting with ZigBee and Z-Wave devices since the two techniques are widely used by IoT devices; specifically, the mediator is built on a Raspberry Pi with a Digi XStick USB dongle (ZB mesh version) and an Aeotec Z-Stick (Gen5) to extend ZigBee and Z-Wave capabilities, respectively.

\subsection{Connecting the Hub and Platform}
To interact with a target platform on behalf of a real device, the mediator creates a virtual device which could: (1) talk with the hub with a communication technique supported by it, and (2) be identified as a compatible device by the platform framework. Most emerging platforms support various connectivity protocols for developers to build customized network devices; for example, SmartThings supports LAN- and cloud-based device integration \cite{lanconnected2018}, openHAB supports Message Queuing Telemetry Transport (MQTT) protocol \cite{mqtt-openhab2019}, Mozilla IoT provides REST-based Web Things framekwork and APIs \cite{webofthings2019}, and Wink allows creating RESTful API devices \cite{winkrestful2019}. This feature alleviates the workload for interfacing with a target platform. We implement the mediator to work with two representative platforms: SmartThings and openHAB. Due to page limit, we present the openHAB part in Appendix~\ref{appendix_openhab_implementation}.

\begin{figure}[tb]
	\centering
	\includegraphics[width=0.48\textwidth]{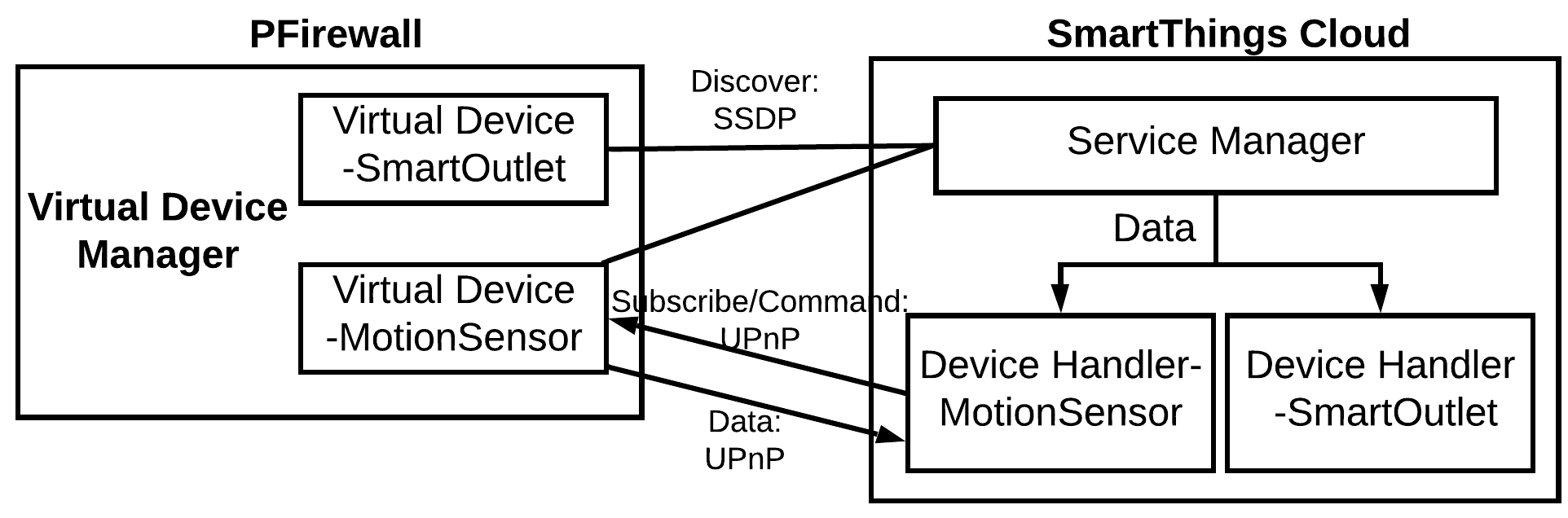}\\
	\caption{Overview of interfacing with SmartThings.}\label{fig_upnp_implementation}
\end{figure}

\vspace{3pt}
\noindent\textbf{Interfacing with SmartThings}\vspace{3pt} \\
We choose LAN as the protocol for communicating with the SmartThings hub since \tool is designed to be segregated from a DMZ by a firewall; thus attackers cannot initiate any connection to \tool to obtain data. SmartThings provides a \texttt{device handler} (see Section~\ref{section_background}) for abstracting each supported device type; accordingly, we build a virtual device (VD) type for each \texttt{device handler} (DH) that originally supports ZigBee or Z-Wave devices, as shown in Fig.~\ref{fig_upnp_implementation}. We develop a \texttt{service manager} SmartApp on SmartThings that uses SSDP (Simple Service Discovery Protocol) to discover VD instances on the LAN. To be considered as different devices (SmartThings uses IP and port to uniquely identify a device), each VD instance is launched on a different port. After discovered a device, the service manager adds it as a \texttt{child device}. When a \texttt{child device} is added, SmartThings automatically selects a DH to abstract it according to the \texttt{model} property of the \texttt{child device}; thus, we make the \texttt{model} property of the VD instance, the \texttt{child device} the same as the \texttt{name} of the target DH that is used to represent the corresponding real device. After the initial connection, a VD instance on the mediator side interacts with a DH instance on the SmartThings with the UPnP (Universal Plug and Play) protocol, which uses SOAP (Simple Object Access Protocol) messages. Additionally, we adapting all DHs for ZigBee/Z-Wave devices available in SmartThings IDE. In each DH, we add a \texttt{subscribe()} function which accomplishes the SUBSCRIBE step for UPnP communication; when a DH is instantiated (which means a VD instance is created and a child device is added), it uses the IP and port to send a SUBSCRIBE SOAP message to the VD instance, providing its IP and port information. Moreover, we change the code in \texttt{parse} and command-related functions for receiving ZigBee/Z-Wave data and sending ZigBee/Z-Wave commands respectively, to code for receving and sending SOAP messages in each DH. Thus, the VD and DH instances become addressable to each other and realize a subscribe/publish based UPnP communication to report data and send commands. 


\section{Evaluation} 
\subsection{Evaluation Setup}
\begin{figure}[tb]
  \centering
  \subfigure[The office]{
    \label{fig_lab} 
    \includegraphics[width=0.23\textwidth]{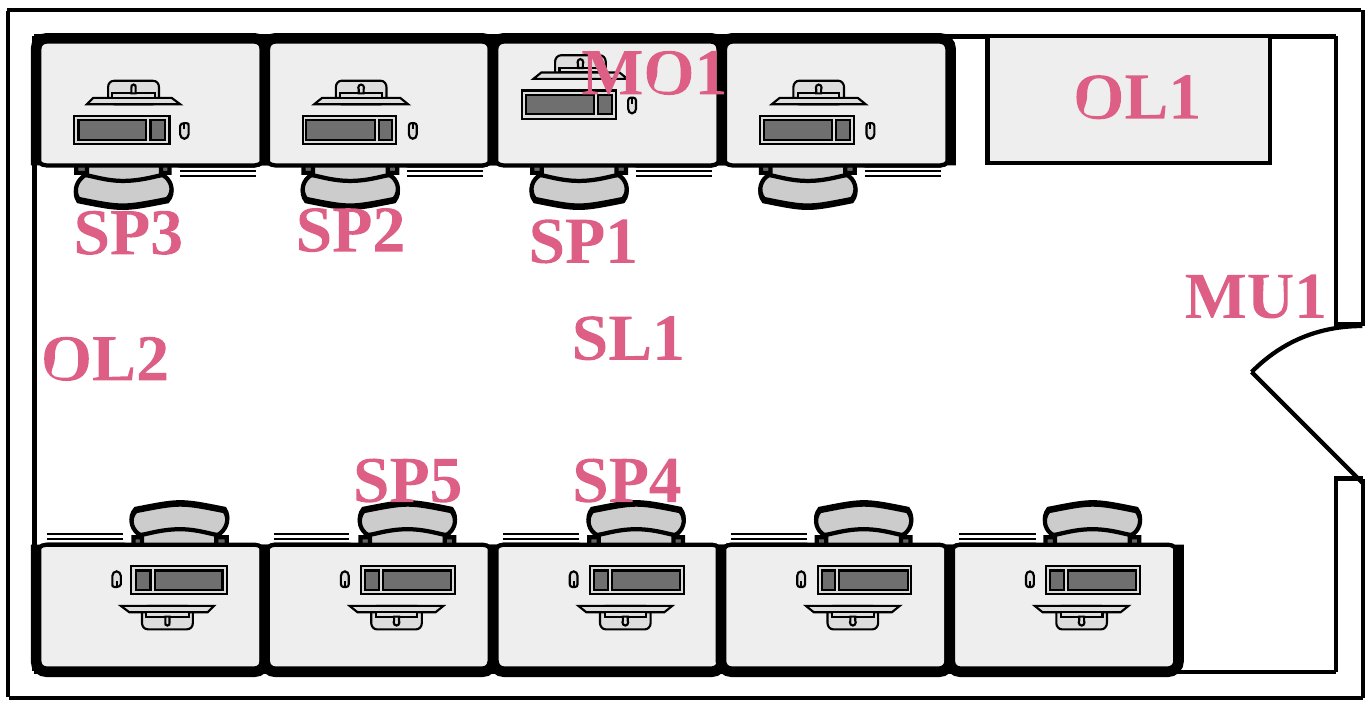}}
  \subfigure[The apartment]{
    \label{fig_home} 
    \includegraphics[width=0.225\textwidth]{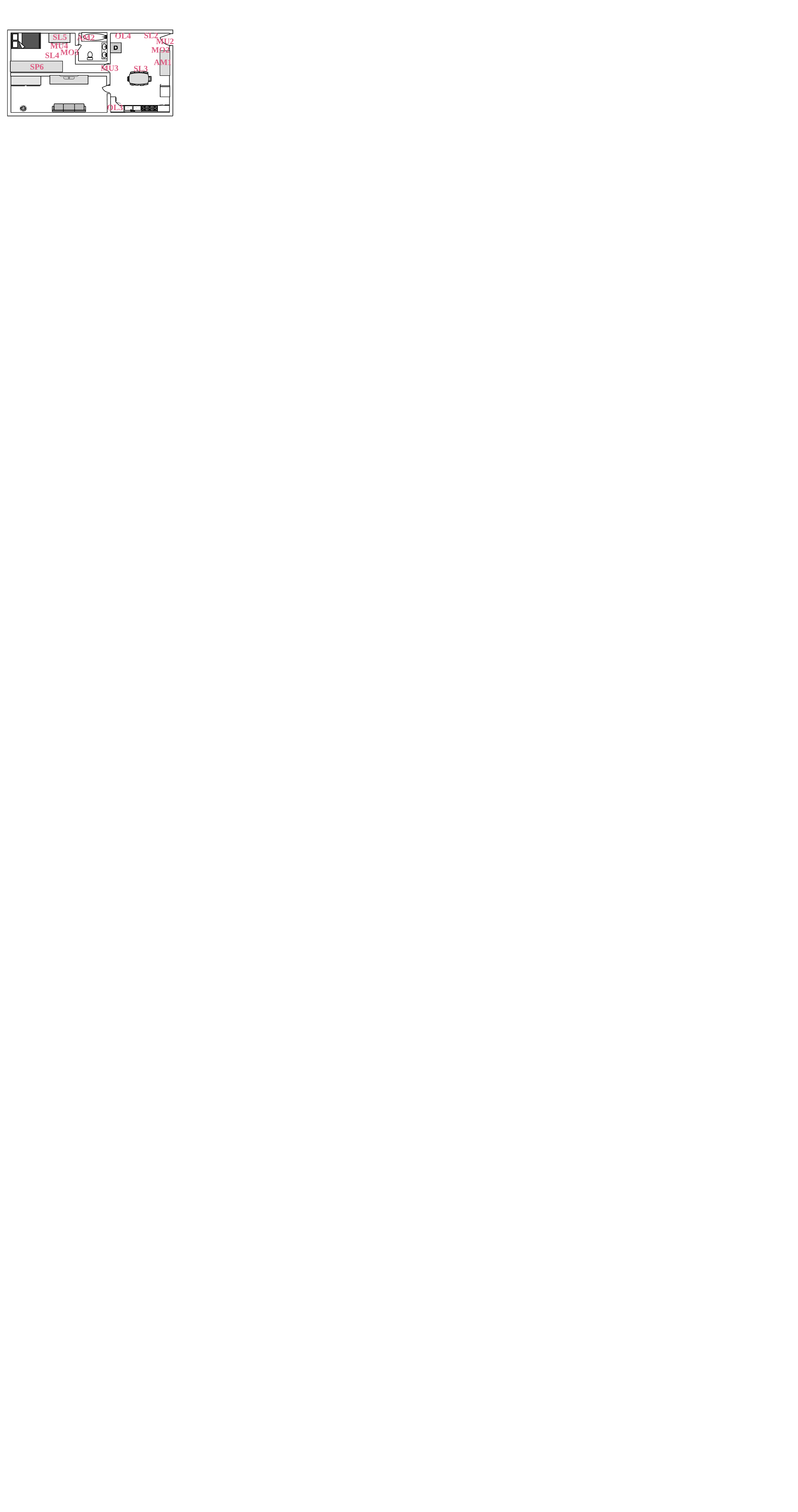}}
  \caption{The layout and device placement in the two testbeds.}
  \label{fig_setup} 
  \vspace{-2.5mm}
\end{figure}

We build two real-world testbeds for evaluating the performance of \tool: an office with 5 members (T$_{1}$) and a two-bedroom apartment with 1 member (T$_{2}$), as shown in Fig.~\ref{fig_setup}. In each testbed (T$_{1}$ and T$_{2}$), we deployed two parallel systems (\texttt{SYS1} and \texttt{SYS2}) by placing two same devices at each position in Fig.~\ref{fig_setup}; \texttt{SYS1} and \texttt{SYS2} have the same device types, numbers, placement and app configuration, as shown in Table~\ref{table_devices}, Fig.~\ref{fig_setup} and Table~\ref{table_rules}. The only difference is that \texttt{SYS1} is a standard SmartThings deployment but \texttt{SYS2} introduces \tool. We bind {\tt SYS1} and {\tt SYS2} in each testbed to two different SmartThings accounts and run them simultaneously but independently. We choose SmartThings in the real-world testbeds because SmartThings provides official apps in its app store, while openHAB needs users to write automation apps and provides no market apps. Instead, we perform some micro-benchmark tests for evaluating openHAB
(see Appendix~\ref{appendix_openhab_evaluation}).

\begin{table}[tb]
	\vspace{-1.5mm}
	\centering
	\scriptsize
	\caption{Devices in the two real-world testbeds}
	\renewcommand\arraystretch{1.1}
	\newcommand{\tabincell}[2]{\begin{tabular}{@{}#1@{}}#2\end{tabular}}
	\newcolumntype{P}[1]{>{\raggedright\arraybackslash}p{#1}}
	\newcolumntype{M}[1]{>{\centering\arraybackslash}m{#1}}
	\begin{tabular}{|M{0.9cm} |P{2.75cm}| P{2.5cm}|c|}	
	\hline
		\bf Testbed & \bf Device ({\tt Abbreviation})  & \centering\bf Attribute & \bf Number\\\hline	 
		\multirow{6}*{\tabincell{c}{Office\\(T$_{1}$)}} 
		~ &	SmartThings hub v2 ({\tt HUB}) &  -- & 1\\ 
		~ & Multipurpose sensor ({\tt MU}) & contact, temperature & 1 \\
		~ & Motion sensor ({\tt MO}) & motion, temperature & 1 \\
		~ & Smart outlet ({\tt OL}) & switch, power & 2 \\
		~ & Smart bulb ({\tt SL}) & switch & 1  \\
		~ & Smartphone ({\tt SP}) & presence & 5\\
		\hline
		\multirow{9}*{\tabincell{c}{Apartment\\(T$_{2}$)}} 
		&	SmartThings hub v2 ({\tt HUB}) & --  & 1 \\
		~ & Multipurpose sensor ({\tt MU}) & contact, temperature & 3\\
		~ & Motion sensor ({\tt MO}) & motion, temperature & 2\\
		~ & Smart outlet ({\tt OL}) & switch, power & 2 \\
		~ & Smart bulb ({\tt SL}) & switch & 4  \\
		~ & Aeotec MultiSensor ({\tt AM}) & motion, humidity, illuminance & 2  \\
		~ & Smartphone ({\tt SP}) & presence & 1\\
	\hline
	\end{tabular}
	\label{table_devices}
\end{table} 

\begin{table*}[tb]
	\vspace{-1.5mm}
	\centering
	\scriptsize
	\caption{SmartApp and device settings for the evaluation environments. O: official app, C: custom app.}
	\renewcommand\arraystretch{1}
	\newcommand{\tabincell}[2]{\begin{tabular}{@{}#1@{}}#2\end{tabular}}
	\newcolumntype{P}[1]{>{\raggedright\arraybackslash}p{#1}}
	\newcolumntype{M}[1]{>{\centering\arraybackslash}m{#1}}
	\begin{tabular}{|M{0.65cm} |P{5.5cm}| P{10.9cm}|}	
	\hline
		\bf Testbed & \bf SmartApps ({\tt Abbreviation})({\tt Source})  & \bf Description and Device Bindings  \\\hline	 
		\multirow{6}*{\tabincell{c}{T$_{1}$}} 
		~ &	UndeadEarlyWarning ({\tt UEW})({\tt O}) & When door ({\tt MU1}) is opened, turn on light ({\tt SL1}).  \\ 
		~ & LightsOffWithNoMotionAndPresence ({\tt LON})({\tt O}) & When no motion ({\tt MO1}) or presence ({\tt SP1$\sim$5}) is detected for 5 minutes, turn off light ({\tt SL1}).    \\
		~ & MyAutoCoffee ({\tt MAC})({\tt C}) & When presence ({\tt SP1}) becomes present, if time is before 12am, turn on coffee machine ({\tt OL1}). \\
		~ & MyAutoHeater ({\tt MAH})({\tt C}) & When motion ({\tt MO1}) detected, if temperature ({\tt MU1}) $< 70^{\circ} F$, turn on heater ({\tt OL2}).   \\
		~ & MyFitnessNotification ({\tt MFN})({\tt C}) & When motion ({\tt MO1}) is active for longer than 60 minutes, send a message to alert.     \\
		~ & StrangerNotification ({\tt STN})({\tt C}) & When door ({\tt MU1}) is open, if no presence ({\tt SP2$\sim$5}), send a message. \\
		\hline
		\multirow{8}*{\tabincell{c}{T$_{2}$}} 
		&	UndeadEarlyWarning ({\tt UEW})({\tt O}) & When door is opened ({\tt MU2}), turn on light ({\tt SL2}).   \\
		~ & SmartLights ({\tt SML})({\tt O}) & When motion ({\tt MO2}) active if illuminance ({\tt AM1}) $< 15$ LUX, turn on light ({\tt SL3}). \\
		~ & TurnOnOnlyIfIArriveAfterSunset ({\tt TOO})({\tt O}) & When presence ({\tt SP6}) becomes present if between 5-8pm, turn on oven ({\tt OL3}). \\
		~ & TextMeWhenThere'sMotionAndI'mNotHere ({\tt TMW})({\tt O}) & When motion ({\tt MO2}) active if not presence ({\tt SP6}), send a notification. \\
        ~ &	LetThereBeLight! ({\tt LTB})({\tt O}) & When wardrobe door ({\tt MU4}) open, turn on light ({\tt SL5}); when door ({\tt MU4}) close, turn off light ({\tt SL5}). \\
		~ & VirtualThermostat ({\tt VIT})({\tt O}) & When motion ({\tt MO2}) is detected if temperature ({\tt MU3}) $<68^{\circ} F$, turn on heater ({\tt OL4}); when motion ({\tt MO2}) inactive for 20 minutes, turn off heater ({\tt OL4}). \\
		~ & SmartBedroomLight ({\tt SBL})({\tt C}) & When door ({\tt MU3}) opened, turn on light ({\tt SL4}); when door ({\tt MU3}) closed if motion ({\tt MO3}) inactive for 5 minutes, turn off light ({\tt SL4}). \\
		~ & NotifyMeWhenSomeoneFaints ({\tt NMW})({\tt C}) & When humidity ({\tt AM2}) exceeds 85\% if motion ({\tt AM2}) active but motion ({\tt MO3}) keeps inactive for 30 minutes, send a notification. \\
	\hline
	\end{tabular}
	\label{table_rules}
\end{table*}  

\subsection{Performance of Data Mediating}
\label{label_performance_mediating}
To test the correctness of \tool mediator, we disable the data filtering in \texttt{SYS2} of both testbeds, i.e., the mediator simply forwards the IoT data to SmartThings without executing policies. To capture received data by SmartThings, we insert \texttt{log.debug} code into the \texttt{parse} methods in all \texttt{device handlers} for the tested devices, which allows us to record the event logs per device on SmartThings web IDE. We observe that there exist duplicate events in the captured SmartThings event logs, so we remove duplicates before analyses; consecutive events that have the same modality (the same device, attribute, value) and very close timestamps (not longer than 1 second) are regarded as the duplicates. We run the above setting in \texttt{SYS2} of both testbeds for 10 days and compare the data sequence of each device received by \tool mediator and SmartThings. Table~\ref{table_mediating_result} shows the total numbers of received data per device and the number of inconsistencies in the data sequences. The result shows that our mediator works effectively and correctly in relaying the received data to the platform.
 
\begin{table}[tb]
	\vspace{-1.5mm}
	\centering
	\scriptsize
	\caption{Statistics of the data received by \tool mediator and that received by SmartThings for the evaluation of data mediating. Due to page limits, we only present the result of one device for each device type. \textbf{Total}: the total data volume received by SmartThings cloud and \tool mediator, respectively.
	}
	\renewcommand\arraystretch{1.1}
	\newcommand{\tabincell}[2]{\begin{tabular}{@{}#1@{}}#2\end{tabular}}
	\newcolumntype{P}[1]{>{\raggedright\arraybackslash}p{#1}}
	\newcolumntype{M}[1]{>{\centering\arraybackslash}m{#1}}
	\newcolumntype{R}[1]{>{\raggedleft\arraybackslash}p{#1}}
	\begin{tabular}{|M{1cm} |c| M{1.5cm}|M{1.5cm}|M{1.6cm}|}	
	\hline
		\bf Testbed & \centering\bf Device  & \bf Attribute  & \bf Total & \bf Inconsistency \\\hline	 
		\multirow{7}*{\tabincell{c}{T$_{1}$}} 
		& {\tt MU1} & contact & 1960, 1960 & 0 \\
		~ & {\tt MU1} & temperature & 174, 174 & 0 \\
		~ & {\tt MO1} & motion & 2198, 2198 & 0\\
		~ & {\tt MO1} & temperature & 325, 325 & 0\\
		~ & {\tt OL1} & switch & 38, 38 & 0\\
		~ & {\tt SL1} & switch & 24, 24 & 0 \\
		~ & {\tt SP1} & presence & 62, 62 & 0 \\
		\hline
		\multirow{3}*{\tabincell{c}{T$_{2}$}} 
		 & {\tt AM1} & motion & 384, 384 & 0\\  
		~ & {\tt AM1} & humidity & 656 , 656  & 0\\
		~ & {\tt AM1} & illumance & 927, 927  & 0\\
	\hline
	\end{tabular}
	\label{table_mediating_result}
\end{table}

\subsection{Performance of Policy System}
To test the performance of our policy system, we establish a comparative experiment by running \texttt{SYS1} and \texttt{SYS2} simultaneously in both testbeds for another 10 days. We enable data filtering in \texttt{SYS2}, so \texttt{SYS2} in this experiment runs the data-minimizaion policies. Also, we define two extra user-specified policies: \textbf{UP1} (DO NOT report {\tt MO1.motion} data between 5pm to 10pm) in T$_{1}$ and \textbf{UP2} (DO NOT report {\tt MU2.contact} data between 8am to 6pm) in T$_{1}$.

\begin{table}[tb]
	\vspace{-1.5mm}
	\centering
	\scriptsize
	\caption{Statistics of SmartApp method call logs. $MC$: the number of method calls; $INC$: the number of inconsistencies; $INCA$: the number of inconsistencies after eliminating redundant method calls.
	}
	\renewcommand\arraystretch{1.1}
	\newcommand{\tabincell}[2]{\begin{tabular}{@{}#1@{}}#2\end{tabular}}
	\newcolumntype{P}[1]{>{\raggedright\arraybackslash}p{#1}}
	\newcolumntype{M}[1]{>{\centering\arraybackslash}m{#1}}
	\newcolumntype{R}[1]{>{\raggedleft\arraybackslash}p{#1}}
	\begin{tabular}{|M{1cm} |M{0.6cm}|M{1.2cm}|M{1.2cm}|M{0.8cm}|M{0.8cm}|}	
	\hline
		\multirow{2}{*}{\tabincell{c}{\textbf{Testbed}}} & \multirow{2}{*}{\tabincell{c}{\textbf{App}}} 
		 & \multicolumn{4}{c|}{\textbf{Data Control Results}}  \\ 
		\cline{3-6}
		& & \tabincell{c}{$MC$ in \texttt{SYS1}}  & \tabincell{c}{$MC$ in \texttt{SYS2}}  & \textbf{\tabincell{c}{$INC$}} & \textbf{\tabincell{c}{$INCA$}}\\ \hline
	
		\multirow{6}*{\tabincell{c}{T$_{1}$}} 
		& {\tt UEW}                 &  971 & 11 & 960   &  0\\
		~ & {\tt LON}                  &  11 & 11 &  0        &  0   \\
		~ & {\tt MAC}                 &  9 & 12    & 3      &  1\\
		~ & {\tt MAH}              &  1059 & 47 &  1012   &  0\\
		~ & {\tt MFN}                   &  13 & 11  & 2       &  2\\
		~ & {\tt STN}                    &  5 & 5      & 0     &  0\\
		\hline
		\multirow{8}*{\tabincell{c}{T$_{2}$}} 
		 & {\tt UEW}                   &  26& 10  & 16      &   3\\
		~ & {\tt SML}                 &  41 & 41   &  0     &   0\\
		~ & {\tt TOO}                  &  11 &  7  &  4      &   0\\
		~ & {\tt TMW}                    &  3  & 3    &  0      &   0\\
		~ & {\tt LTB}                 &  42 & 42    & 0      &   0\\
		~ & {\tt VIT}                 &  235 & 49 &  186    &   0\\
		~ & {\tt SBL}                &  164 & 55 &  109     &   0\\
		~ & {\tt NMW}                   &  3 & 3    & 0       &   0\\
	\hline
	\end{tabular}
	\label{table_mediating_result_apps}
\end{table}

\subsubsection{Correctness and Reliability}
\label{label_correctness_policy}
Comparing the received data sequences is meaningless since data are filtered in \texttt{SYS2}, so we test the correctness of the execution of SmartApps. To capture the execution of apps, we manually insert logging code into the installed SmartApps to record the method calls for controlling devices and sending notifications. We compare the method call sequences of each app in {\tt SYS1} and {\tt SYS2} and calculate the number of inconsistencies. We summarize the result in Table~\ref{table_mediating_result_apps}. We figure that the $INC$ values of some apps are large. This is because SmartThings apps do not check a device's current status before sending it a command and thus redundant method calls are made
while \tool in design disables redundant automation commands to reduce reporting data.
For instance, the app \texttt{UEW} calls the light turn-on method every time the door ({\tt MU1}) is opened, no matter the light is ``on'' or ``off''; however, the redundant method calls are avoided by our data flow policies if the light's status is already ``on''.
Thus, inconsistencies are detected in some apps. To eliminate the impact of redundant automation on the evaluation of automation accuracy, we capture and remove the redundant method calls from method call sequences in \texttt{SYS1} by analyzing app and device logs; specifically, if a method call's effect is to change a device to a state the device is already in, this method call is identified as redundant and removed from the sequence. We recalculate the inconsistencies, denoted as $INCA$. As show in Table~\ref{table_mediating_result_apps}, $INCA$ in most apps are 0 except in four apps: {\tt MAC}, {\tt MFN} in T$_{1}$ and {\tt UEW} in T$_{2}$. We manually analyze the causes of these inconsistencies by examining the device event and method call logs. 
we find that the event log of \texttt{SP1} in \texttt{SYS2} has one more ``present'' than that of {\tt SYS1}. This is because SmartThings detects presence by monitoring the distance of a smartphone (GPS data) from the in-home hub while \tool scans the home WiFi network to examine if a smartphone enters/leaves; when \texttt{SP1} moves around, different presence statuses are detected by the two methods due to distinct detection ranges, leading to the inconsistency in \texttt{MAC}. The inconsistencies in {\tt MFN} and {\tt UEW} appear because user specified policies \textbf{UP1} and \textbf{UP2} block {\tt MO1.motion} and {\tt MU2.contact} data during certain periods, respectively. We verify that the 2 inconsistencies in {\tt MFN} occur during 5pm-10pm and the 3 inconsistencies in {\tt UEW} occur during 8am-6pm. We also observe that no {\tt MO1.motion} or {\tt MU2.contact} data are received by SmartThings in \texttt{SYS2} during the specified periods in \textbf{UP1} and \textbf{UP2}, respectively. The above result shows the correctness of our policy-based data flow control in enforcing user-specified policies and in preserving home automation functionalities by generating data-minimization policies.  

\subsubsection{Latency}
\begin{figure}[tb] 
	\centering
	\includegraphics[width=0.40\textwidth]{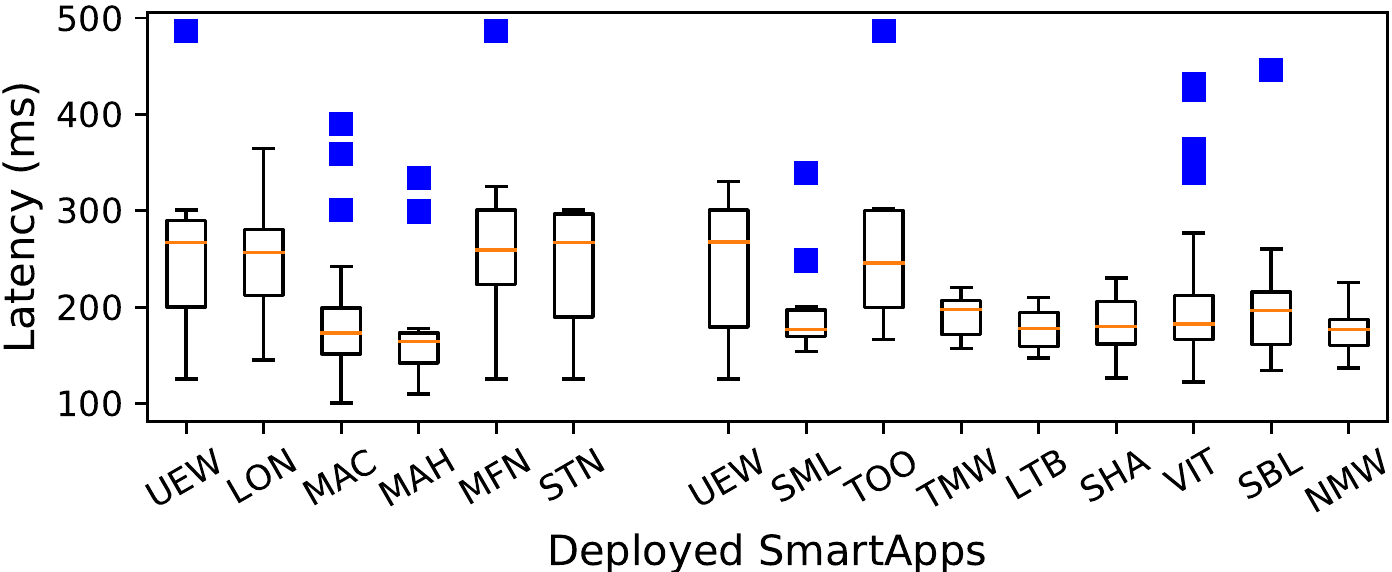}\\ 
	\caption{Automation latency introduced by \tool. The boxes show the maximum, quartile, averaged and minimum values of the majority latencies per app. The blue squares are some outliers.}\label{fig_automation_latency} 
\end{figure}

We show the efficiency of \tool by testing the introduced automation latency (mediating delay plus policy execution delay). We obtain the result by computing the timestamp difference of the same command in both command sequences ({\tt SYS1} and {\tt SYS2}). We exclude the outliers from our calculation where the command in {\tt SYS1} is even issued after {\tt SYS2} to reduce the influence of network delay and the cloud response latency on the result. We calculate the automation latency for each SmartApp in both testbeds and show the result in Figure~\ref{fig_automation_latency}. The automation latency ranges from 124.7 to 486.4 millisecond. An averaged latency of 210.6 millisecond is a tradeoff for using \tool to mitigate privacy leakage, although the latency is completely acceptable for most automation apps. 

\subsubsection{Reduction of Data Leakage}
\label{section_data_reduction}
To show the effectiveness of data filtering, we compare the data volume reported by each device in the {\tt SYS1} and {\tt SYS2} of both testbeds. As show in Table~\ref{table_reduction_result}, \tool blocks 96.87\% IoT data on averaged. More than 99\% of float-value sensor readings and device states (i.e., ON/OFF states of coffee machines, setpoints of thermostats, locked/unlocked states of smart locks, etc.); thus, \tool prevents the smart home platforms and potential attackers from learning the private information of smart homes and homeowners based on float-value sensors and household appliances. \tool also reduces the reporting of binary-value sensor attributes (contact, motion, presence) to distinct extents, according to the specific automation app semantics and app-device bindings. The relative reduction rate $RR$ of binary-value attributes are smaller than float-value attributes in general, since binary attributes are used for triggering the execution of automation apps in most cases and hence cannot be totally blocked. 

\begin{table}[tb]
	\vspace{-1.5mm}
	\centering
	\scriptsize
	\caption{Comparison of reported data volume per device before and after the deployment of \tool. $VOL$: volume of reported data in {\tt SYS1} and {\tt SYS2}, respectively; $RR$: data reduction rate. We present the result for partial devices. See Appendix~\ref{appendix_complete_data_volume_comparison} for the result of all deployed devices.
	}
	\renewcommand\arraystretch{1.1}
	\newcommand{\tabincell}[2]{\begin{tabular}{@{}#1@{}}#2\end{tabular}}
	\newcolumntype{P}[1]{>{\raggedright\arraybackslash}p{#1}}
	\newcolumntype{M}[1]{>{\centering\arraybackslash}m{#1}}
	\newcolumntype{R}[1]{>{\raggedleft\arraybackslash}p{#1}}
	\begin{tabular}{M{0.4cm}M{1.05cm}M{1cm}M{0.4cm}M{1.05cm}M{1cm}M{0.4cm}}	
	\toprule
    \textbf{Dev} & \textbf{Attr} & $VOL$ & $RR$  & \textbf{Attr} & $VOL$       & $RR$ \\
    \midrule
    {\tt MU1}   &  contact  & 1924, 22  & 0.98  &  temperature  & 142, 6  & 0.96 \\ 
	{\tt MO1}   &  motion   & 2266, 47  & 0.98  &  temperature  & 307, 0  & 1 \\ 
	{\tt OL1}   &  switch   &  29, 0  & 1  &    &   &  \\
	{\tt SL1}   &  switch   &  22, 0  & 1  &    &   &  \\
	{\tt SP1}   &  presence  & 34, 24  & 0.29  &    &   &  \\
	{\tt MU2}   &  contact  & 52, 24  & 0.54  &  temperature  &  118, 0 & 1 \\
	{\tt MO2}   &  motion  & 364, 68  & 0.81  &  temperature  &  173, 0 & 1 \\
	{\tt OL3}   &  switch  & 44, 0  & 1  &    &   &  \\
	{\tt SL2}   &  switch  & 60, 0  & 1  &    &   &  \\
	{\tt AM1}   &  motion  & 364, 0  & 1 
	\\
	{\tt AM1}   &  illuminance  & 1039, 1  & 0.99  &  humidity  & 668, 0  & 1 \\
	{\tt SP6}   &  presence  & 28, 12  & 0.57  &    &   & \\
	\bottomrule
	\end{tabular}
	\label{table_reduction_result}
\end{table}

\subsubsection{Privacy Gain}
\label{privacy_gain}
To show how privacy preservation is achieved by the reducing data leakage, we compare the potential privacy leakage under several inference attacks with and without \tool.

\begin{figure}[t]
  \centering
  \subfigure[Without data flow control]{
    \label{fig_presence_without} 
    \includegraphics[width=0.23\textwidth]{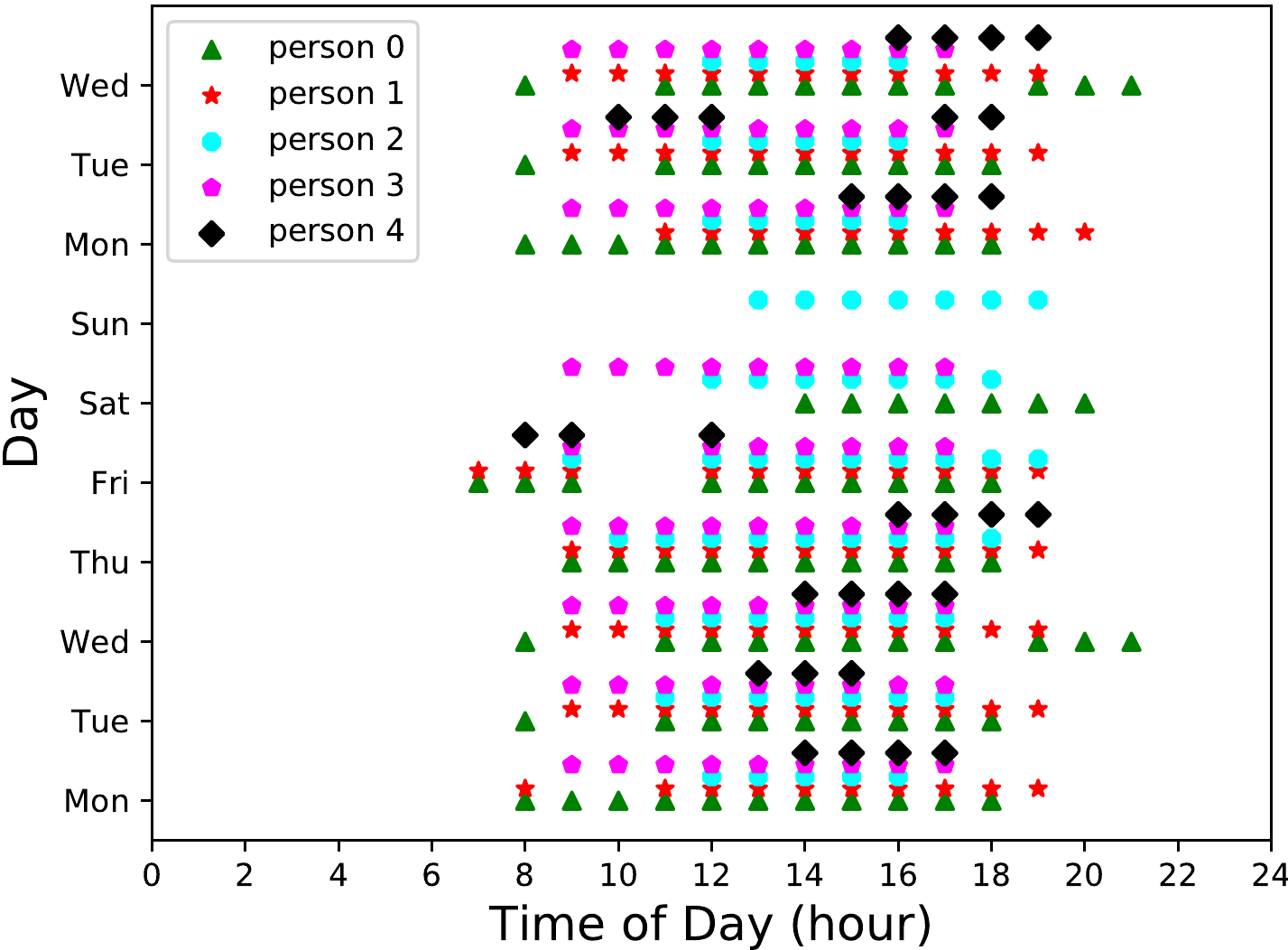}}
  \subfigure[With data flow control]{
    \label{fig_presence_with}
    \includegraphics[width=0.23\textwidth]{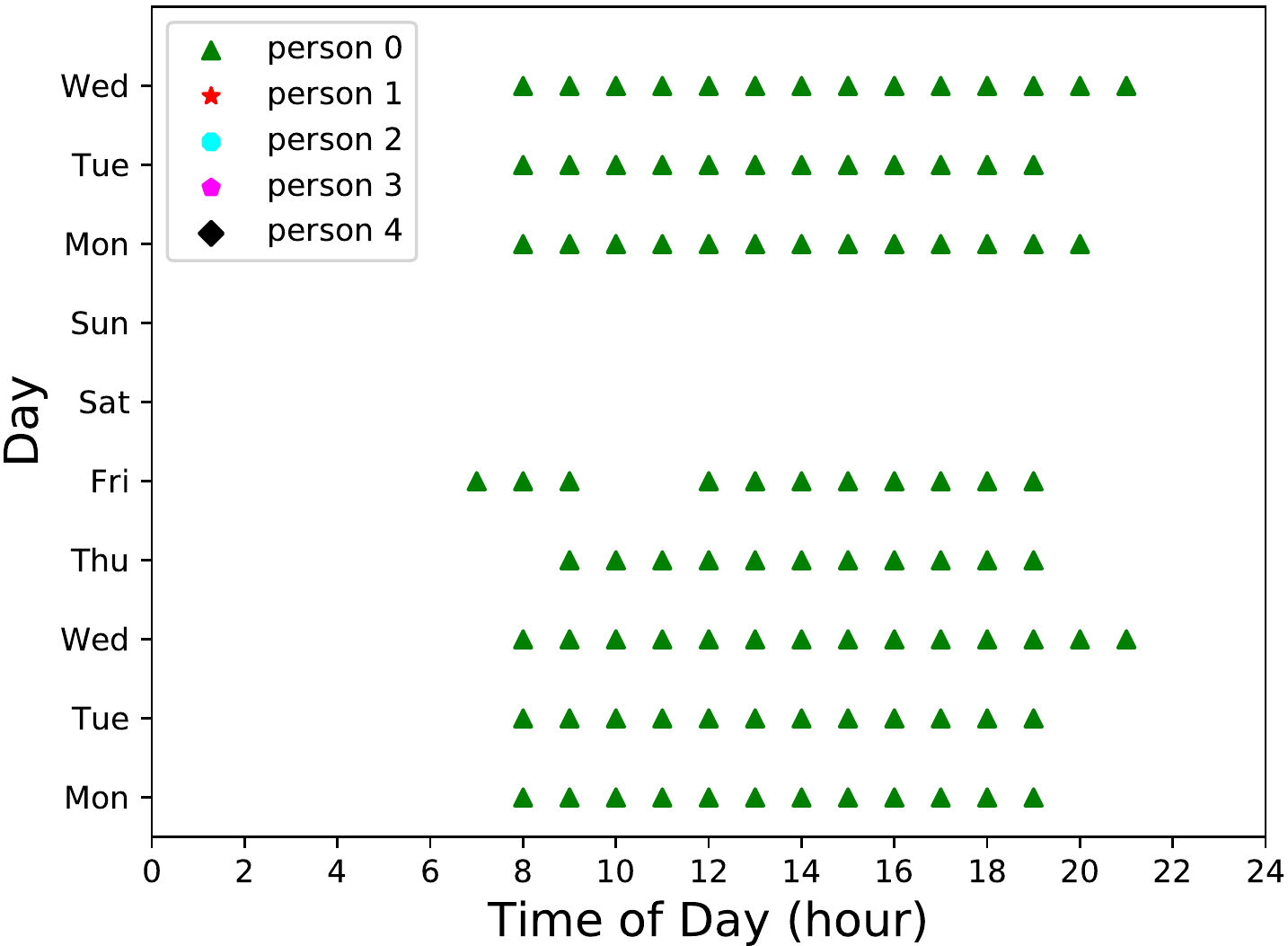}}
  \caption{Inferred user working hours within 10 days with and without data flow control in testbed T$_{1}$. For simplicity of illustration, we round all presence data timestamps to the nearest hours.}
  \label{fig_presence_examples}
\end{figure}

\vspace{3pt}\noindent
\textbf{Office members and events profiling.} By analyzing the presence sensor ({\tt SP1$\sim$5}) data in the research lab testbed (T$_{1}$), the working hours of 5 members (person 1$\sim$5) each of whom carries a presence sensor could be learned, based on their entering and leaving time, as shown in Fig.~\ref{fig_presence_without}. In addition to monitoring user presence in real time, the attacker could also learn the personal working preferences and group events. For example, person 1 may leave for classes each Tuesday and Wednesday; person 3 works less hours than person 1 and 2 during weekdays but shows up more on weekends; person 4 has a more regular routine through the weekdays; person 5 works less hours (4 or so) every day and the hours tend to be in the afternoon; moreover, the members may leave for a group meeting on Friday morning. When \tool is deployed, most presence data are filtered since only the ``present'' events before 12am from {\tt SP1} are required to turn on coffee machine outlet (see app {\tt MAC}). The presence sensor data of the other persons are never sent because their values are kept ``not present'' in the platform database and only ``not present'' events from {\tt SP1} are sent in order for the app {\tt LON} to pass its condition checking. when the last person leaves. which hides the real leaving time of person 1.. Therefore, an attacker could only learn when person 1 arrives the lab room  correctly (see Fig.~\ref{fig_presence_with}).

\begin{figure}[t]
  \centering
  \subfigure[motion data without control]{
    \label{fig_bathroom_motion_without} 
    \includegraphics[width=0.23\textwidth]{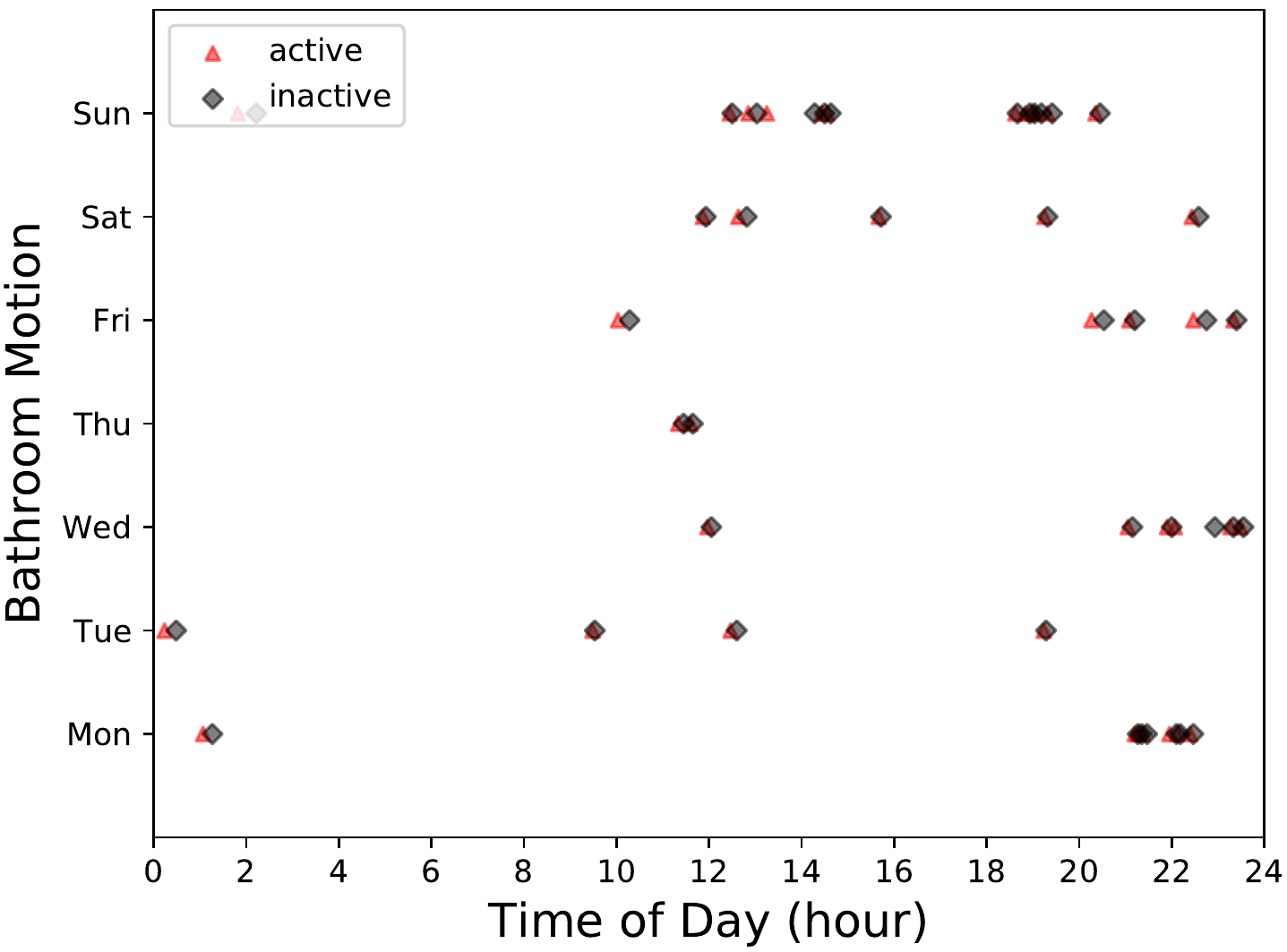}}
  \subfigure[motion data with control]{
    \label{fig_bathroom_motion_with}
    \includegraphics[width=0.23\textwidth]{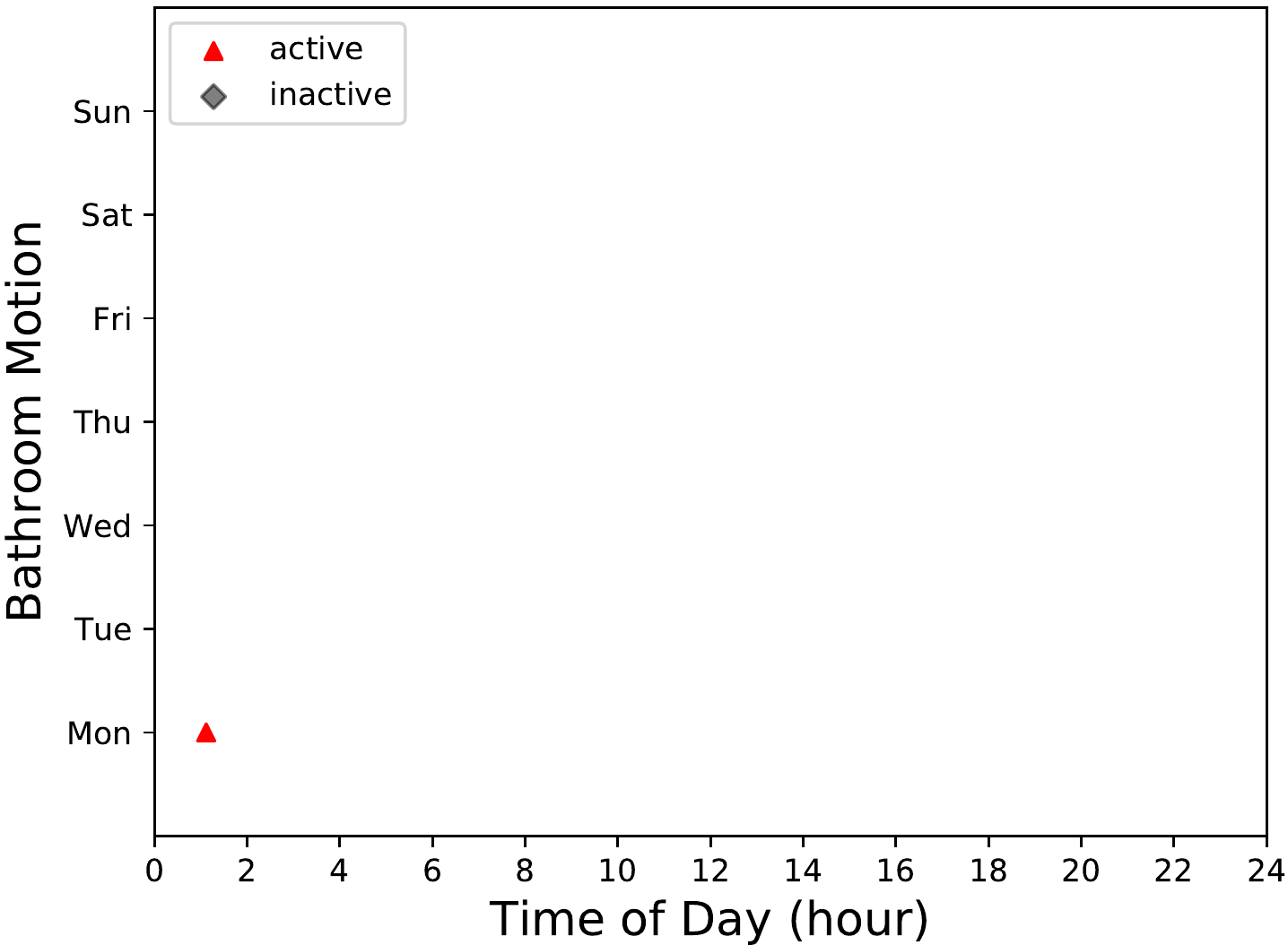}}
  \\
    \subfigure[humidity data without control]{
    \label{fig_bathroom_humidity_without} 
    \includegraphics[width=0.23\textwidth]{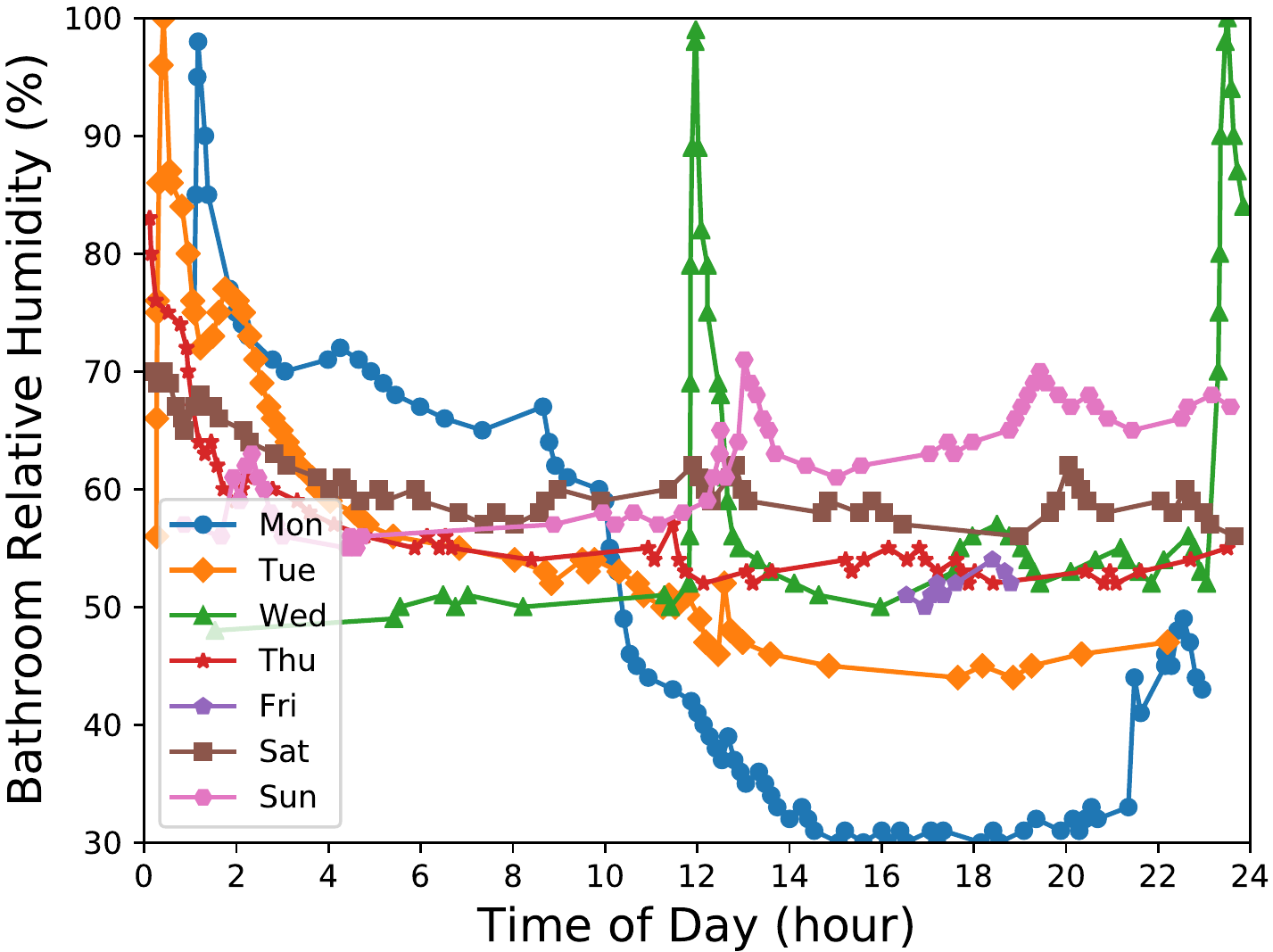}}
  \subfigure[humidity data with control]{
    \label{fig_bathroom_humidity_with}
    \includegraphics[width=0.23\textwidth]{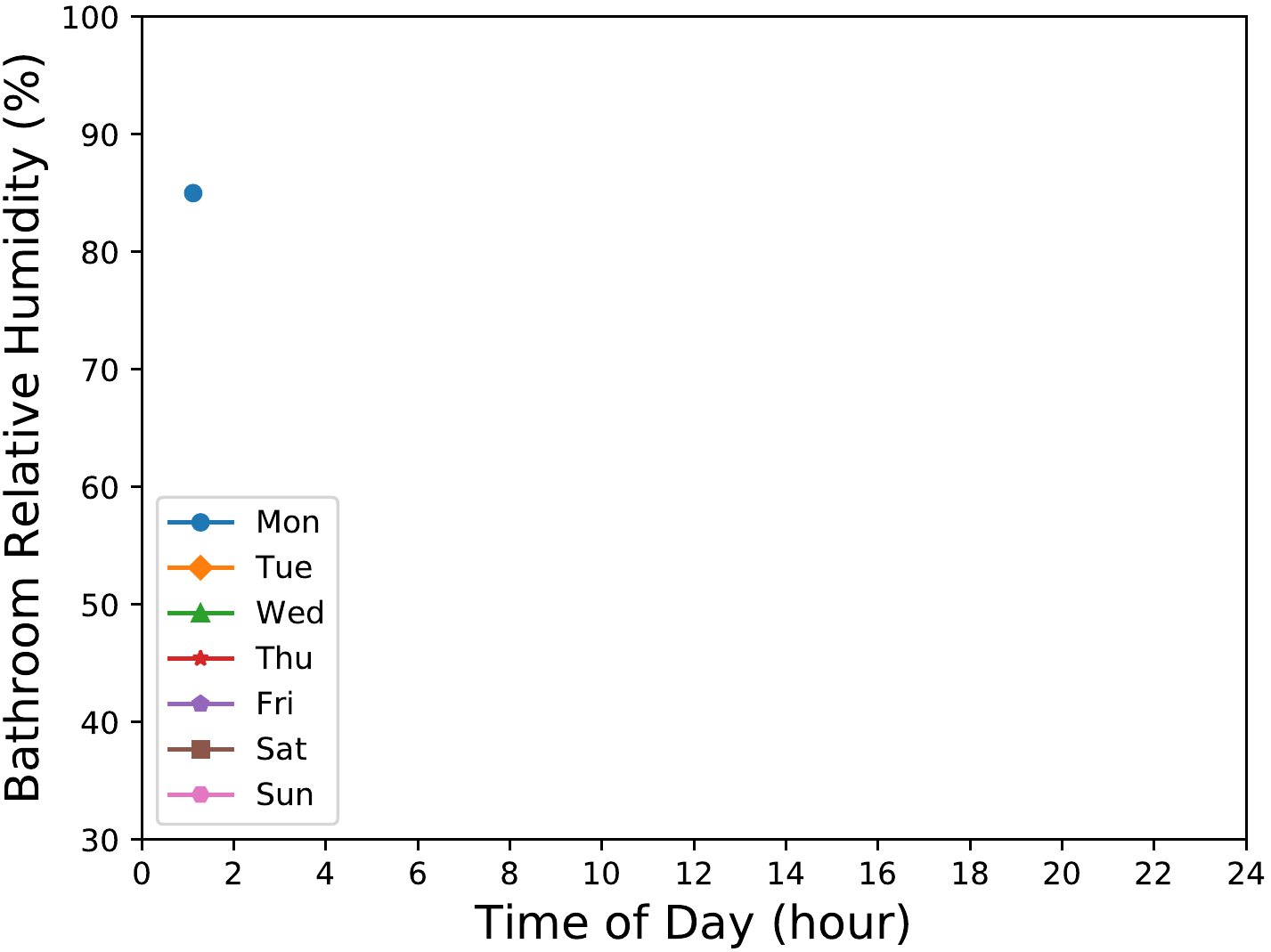}}
    \caption{1-week motion and humidity data in the bathroom of testbed T$_{2}$ received by the platform with and without data flow control. For an clearer display, motion data that indicate shorter than 3-minute bathroom activities are omitted in (a).}
  \label{fig_bathroom_humidity}
\end{figure}

\vspace{3pt}\noindent
\textbf{Bathroom usage monitoring.} By accessing the motion and humidity data of the Aeotec Multisensor ({\tt AM2}) in the apartment testbed (T$_{2}$), an attacker can learn the bathroom usage habits. As depicted in Fig.~\ref{fig_bathroom_motion_without}, the attacker simply combines each ``active'' with the next ``inactive'' event to obtain the start and end time of a bathroom usage. Moreover, the attacker can also use the humidity data (see Fig.~\ref{fig_bathroom_humidity_without}) as additional information to help recognize ``having shower'' activities in the bathroom. In the experiment, the attacker identifies 4 ``having shower'' activities by comparing the humidity values with a common sense threshold (i.e., 85\%). When \tool is applied, the humidity data is rarely sent (for executing the anomaly activity detection app {\tt NMW}) and motion ``active'' ({\tt AM2}) is reported only once to keep the motion value ``active'' in the platform database. As shown in Fig.~\ref{fig_bathroom_motion_with} and \ref{fig_bathroom_humidity_with}, the humidity and motion data are respectively sent only once in our one-week experiment, preventing the attacker from monitoring and learning the bathroom usage habits.   

\begin{figure}[t]
  \centering
  \subfigure[Raw data]{
    \label{fig_raw_power} 
    \includegraphics[width=0.23\textwidth]{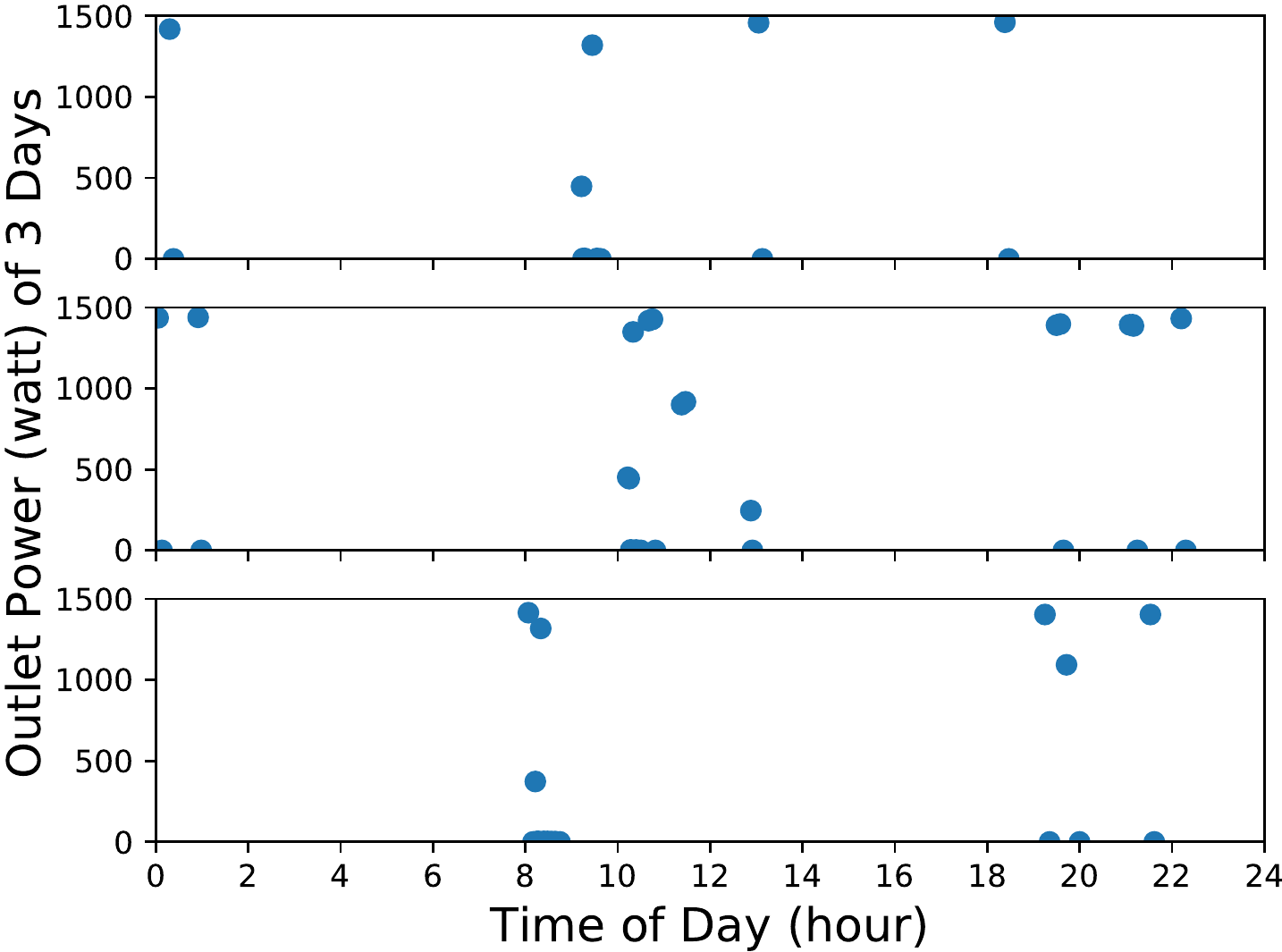}}
   \subfigure[Slicing]{
     \label{fig_power_slided_data}
     \includegraphics[width=0.23\textwidth]{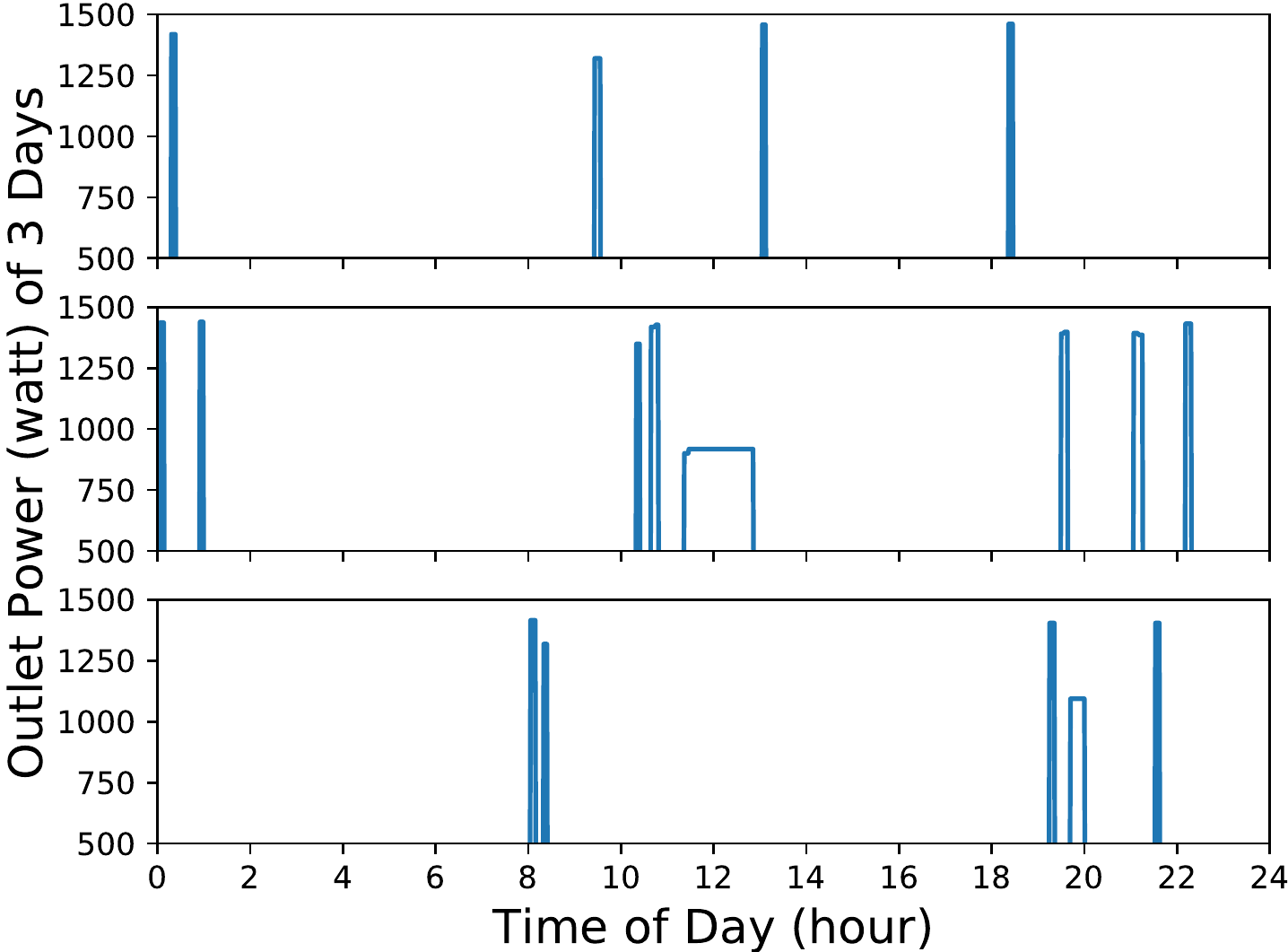}}
     
   \subfigure[K-means Clustering]{
     \label{fig_power_clustering_result}
     \includegraphics[width=0.23\textwidth]{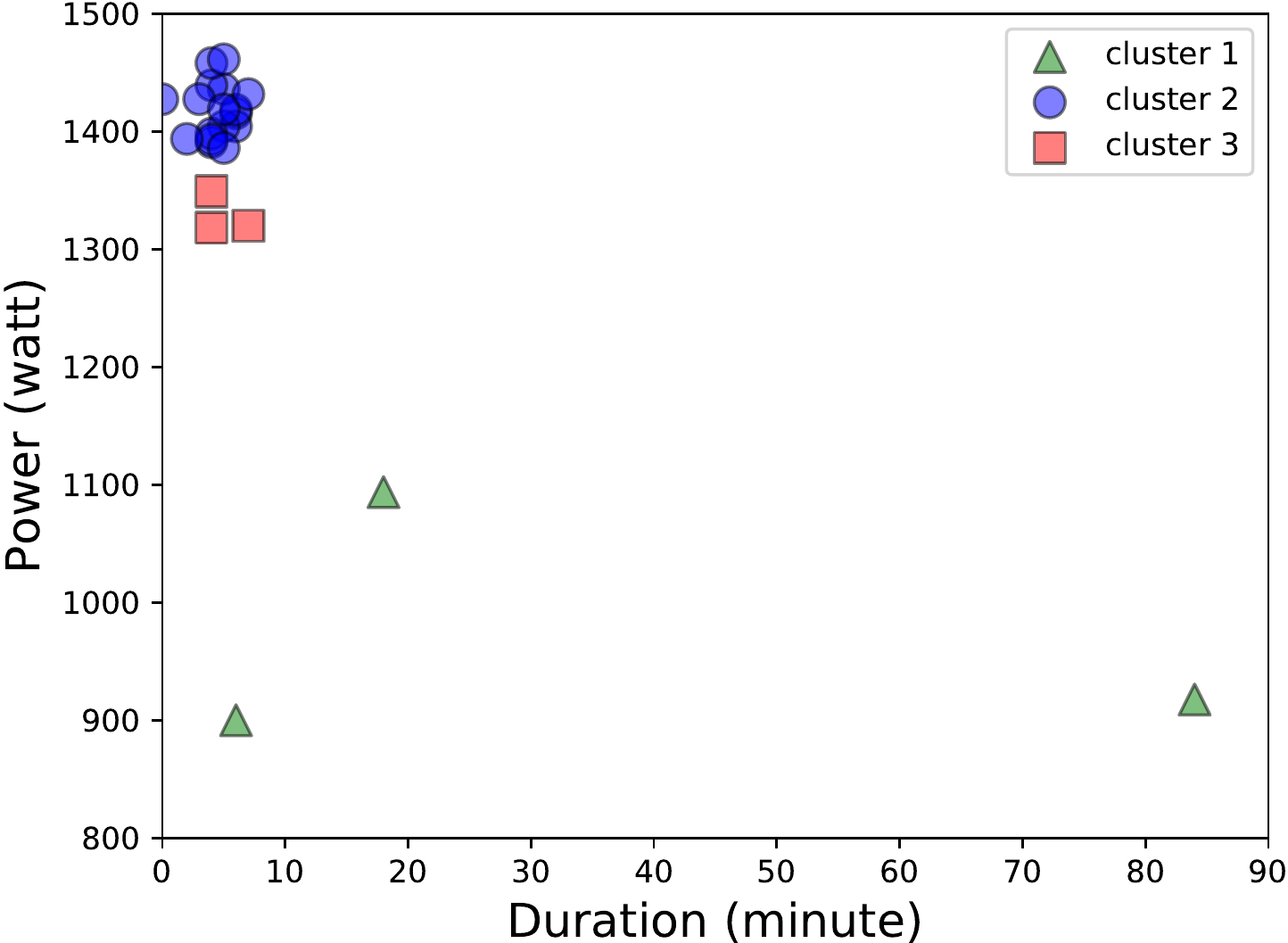}}
    \subfigure[Mapping clusters to appliances]{
    \label{fig_power_mapping_result} 
    \includegraphics[width=0.23\textwidth]{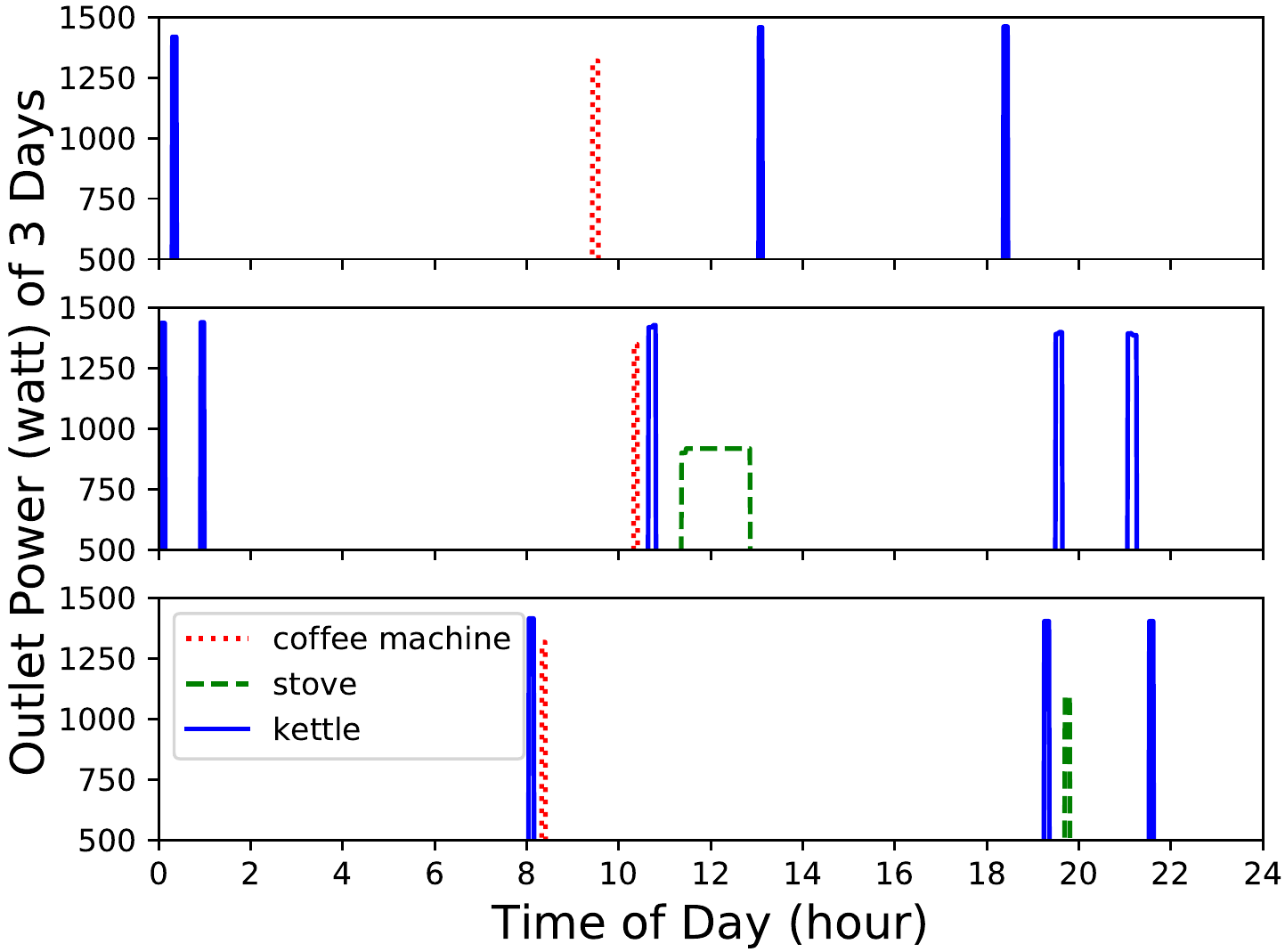}}
  \label{fig_outlet_power}
  \caption{Appliance usage inference over 3-day power data without data flow control.}
\end{figure}

\vspace{3pt}\noindent
\textbf{Appliance monitoring.} Non-intrusive load monitoring (NILM) techniques can infer appliance events based on electricity data, causing privacy concerns \cite{mclaughlin2011protecting, lisovich2008privacy}. We set up another experiment to learn how attackers are prevented from inferring appliance working status and user activities when power data are protected. We connect a microwave, a kettle and a stove to a smart outlet and install an automation app that turns off the outlet when a user leaves home to avoid fire accidents. Although the app only needs a presence sensor data to operate, the outlet also measures real time power data and reports it to outside. To study the incurred privacy risk, we collect the reported raw power data (see Fig.~\ref{fig_raw_power}) for 3 days and perform inference attacks. The attack process includes data pre-processing, clustering and mapping (Fig.~\ref{fig_power_slided_data}-\ref{fig_power_mapping_result}). The inference result achieves 95.7\% precision and 92\% recall in identifying appliance activities when compared with the manually collected ground truth. When \tool operates, all power data are preserved for running this app and hence no user privacy could be inferred from power data. 

\section{Discussion and Limitations}
\label{section_discussion_limitations}
\noindent

\vspace{3pt}
\noindent
\textbf{Can \tool perform home automation and thus get rid of
the cloud?}
Note that \tool has access to all device data and rule semantics from IoT apps. Theoretically, \tool is capable of running a rule engine to execute the extracted semantics; thus, no data is sent to the cloud at all. However, we did not employ this design due to practical considerations. (1) The kick-cloud-out strategy may cause ethical or legal concerns which our research team cannot tackle. The SmartThings cloud can easily verify whether
it is talking with a real SmartThings hub, and cut all the services if not. It means that, while \tool may provide home automation, all other cloud-based services (messaging, storage, and remote management) will be lost. (2) Huge engineering efforts are needed to implement an equivalent rule engine that supports the same programming framework and APIs and maintain them in a long run. 
Therefore, we strategically segregate the data flow control policy engine and the rule engine; \tool only deals with data filtering.

\vspace{3pt}
\noindent
\textbf{User efforts.}
In \tool, users pair IoT devices with the mediator on \tool web interfaces and add the virtual device instances to SmartThings with its companion mobile app; thus, users operation for connecting devices is doubled. We design SmartThings-alike pairing interfaces on the \tool side, which makes pairing on both sides similar and reduces potential confusions. Moreover, we use the browser automation framework Selenium to develop a Python script, which periodically checks the new SmartApps and devices, and installs corresponding instrumented SmartApps (for rule extraction) and custom device handlers (for \tool mediation), respectively. Users only provide their SmartThings accounts to the script and no other operations are required.

\vspace{3pt}
\noindent
\textbf{Generality.}
Although our implementation targets SmartThings and openHAB, the presented approach can be potentially adapted to other ecosystems. As discussed in Section~\ref{section_relay}, it is complete practical to realize a man-in-the-middle mediator in most systems. On one hand, the mediator could be extended to work with as various IoT devices as an open-source platform; on the other, the mediator could interfacing with many platforms via a connectivity technique provided by these platforms for creating and integrating software services and hardware devices as ``things''. Moreover, approaches for extracting automation rules from IoT apps \cite{jia2017contexiot, tian2017smartauth, zhang2018homonit, celik2018soteria, chi2018cross} and mobile/web interfaces \cite{zhang2018homonit, hwang2016data, nguyen2018iotsan} have been broadly studied. We envision that tools are developed by the community for extracting rule semantics from more platforms such that the data-minimization policies can be generated.

\section{Related Work}
\label{section_related_work}

\subsection{Privacy in Smart Home Platforms}
Besides security, privacy is also an important research topic in smart home ecosystems. Zheng et al. \cite{zheng2018user} studied smart home owners' perceptions of privacy risks and actions taken to protect their privacy; the study found that users are unaware of privacy risks from inference algorithms operating on data from their IoT devices, and they expect device manufacturers to protect their privacy though it is not the case. Celik et al. \cite{celik18sensitive} provided a tool for tracking the sensitive data flows in programming frameworks and identified 138 out of 230 apps in SmartThings transmit at least one kind of sensitive data over platform-provided APIs, which means malicious apps have the capability to steal user data collected by the platform. Literature \cite{fernandes2016security} and \cite{jia2017contexiot} also present app-level attacks that can brench user privacy. Closest to our work, FlowFence \cite{fernandes2016flowfence} enforced a data flow control mechanism for sensitive data protection. However, FlowFence protects sensitive data from unauthorized apps rather than the platform, so sensitive data protection still fails to other attacks; FlowFence requires the cooperation from the platforms and app developers to operate.

\subsection{In-hub Security and Privacy Enforcement}
Many in-hub schemes are proposed to enforce security and privacy schemes in the IoT domain. Simpson et al. design a in-hub security manager built atop the smart home hub to patch vulnerable IoT devices and strengthen authentication. The security manager is deployed in a open-source system HomeOS. FACT \cite{lee2017fact} and HanGuard \cite{demetriou2017hanguard} enforce access controls in the middle by implementing controllers on an open-source hub and a programmable WiFi router, respectively. By comparison, these schemes rely on a programmable hub (gateway, router) that can indeed intercept control the communication between home area network and the Internet. However, in cloud-based smart home platforms like SmartThings, communications between the commercial hub and the backend cloud are encrypted \cite{veracodehubresearch} and hence the router can neither decrypt nor modify the packets on demand. \tool controls the communication between IoT devices and the hub in a unified, backward-compatible way, regardless of the specific communication protocol employed by the hub and cloud.

\section{Conclusion}
We presented \tool, a semantics-aware customizable data flow control system
for smart homes, which filters data generated by IoT devices. \tool can automatically generate application-dependent policies based on installed automation apps to block unnecessary data flows and only report the minimum amount of data required for home automation. Furthermore, \tool allows users to customize individual policies according to their own privacy preferences.

We overcame many challenges and designed an elegant man-in-the-middle proxy
based system, which enforces these policies without modifying the platform
or IoT devices. 
We implemented a prototype of \tool and evaluated it in two real-world testbeds. The evaluation results demonstrated that \tool can effectively and efficiently reduce sensitive data leakage without interfering with home automation. It heavily impairs an
attacker's ability to monitor and infer user privacy-sensitive behaviors. In addition to smart homes, the system can 
also significantly enhance privacy protection in many other environments,
such as smart factories and offices, that leverage smart platforms
for IoT device interaction automation and other platform-provided services.

\bibliographystyle{IEEEtran}
\bibliography{references}

\appendix

\begin{figure*}[t]
  \centering
  \subfigure[]{
    \label{subfig_survey_devices}
    \includegraphics[width=0.23\textwidth]{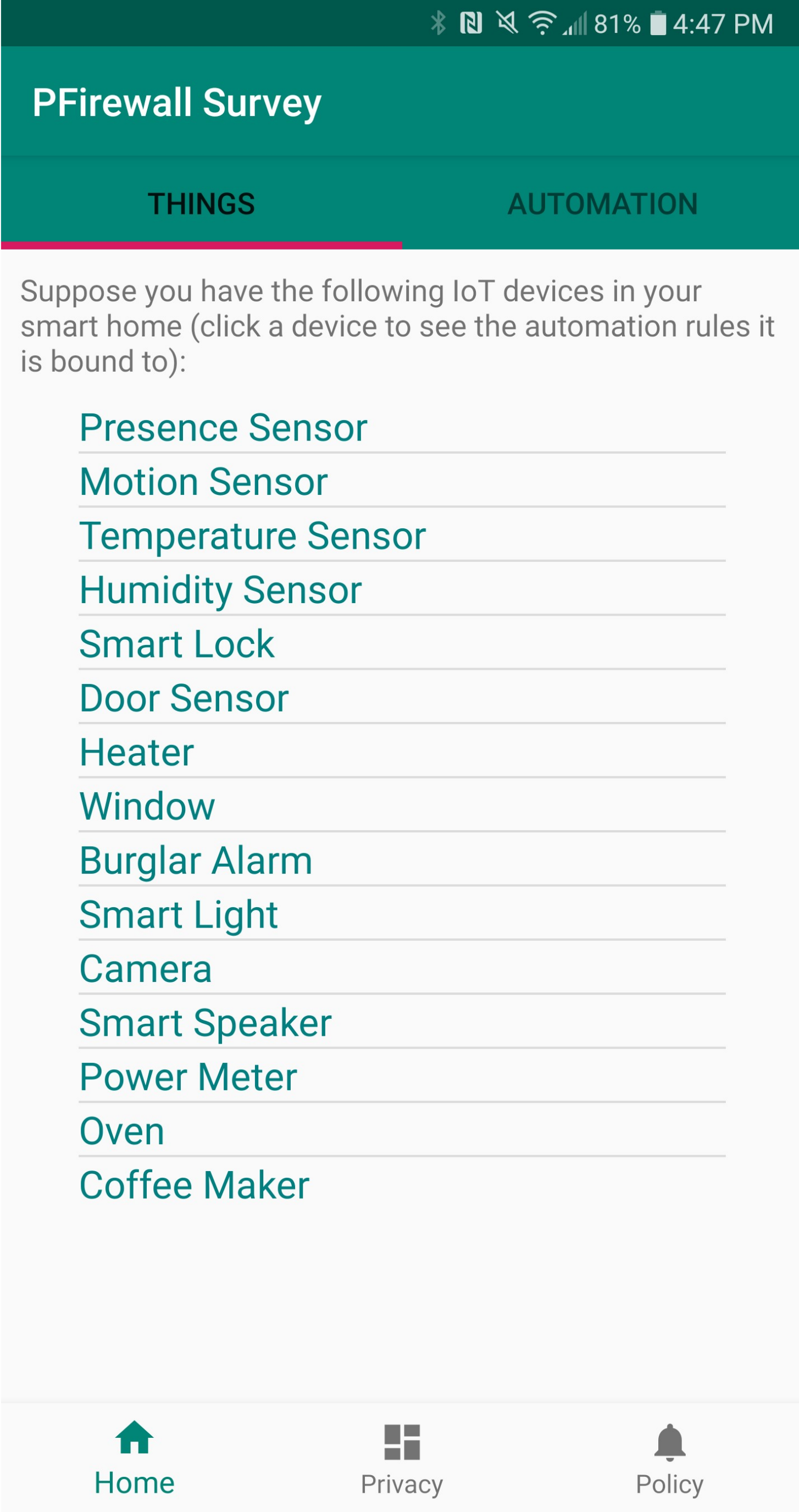}}
   \subfigure[]{
   \label{subfig_survey_automation}
     \includegraphics[width=0.23\textwidth]{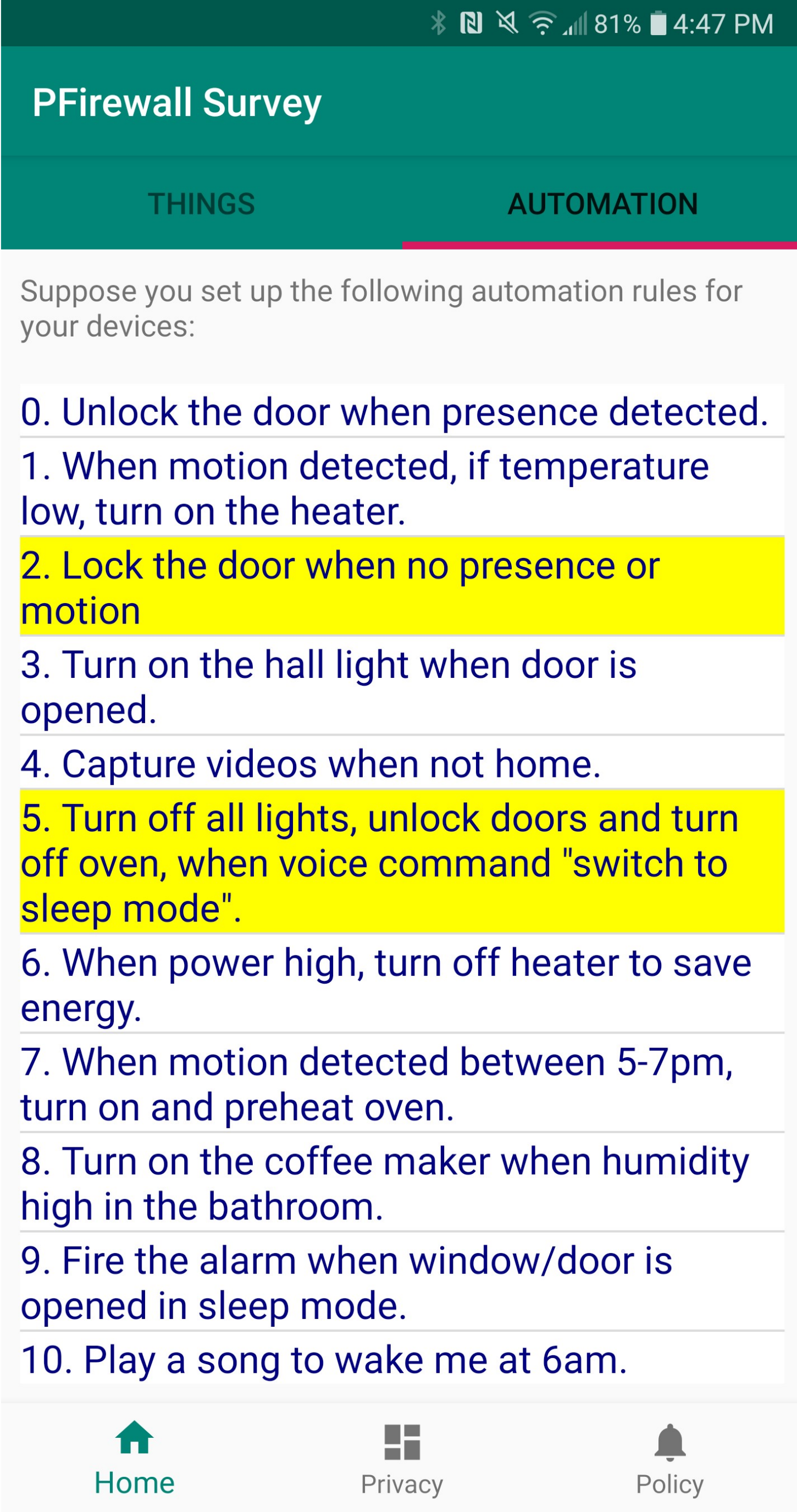}}
     \subfigure[]{
     \label{subfig_survey_tutorial}
    \includegraphics[width=0.23\textwidth]{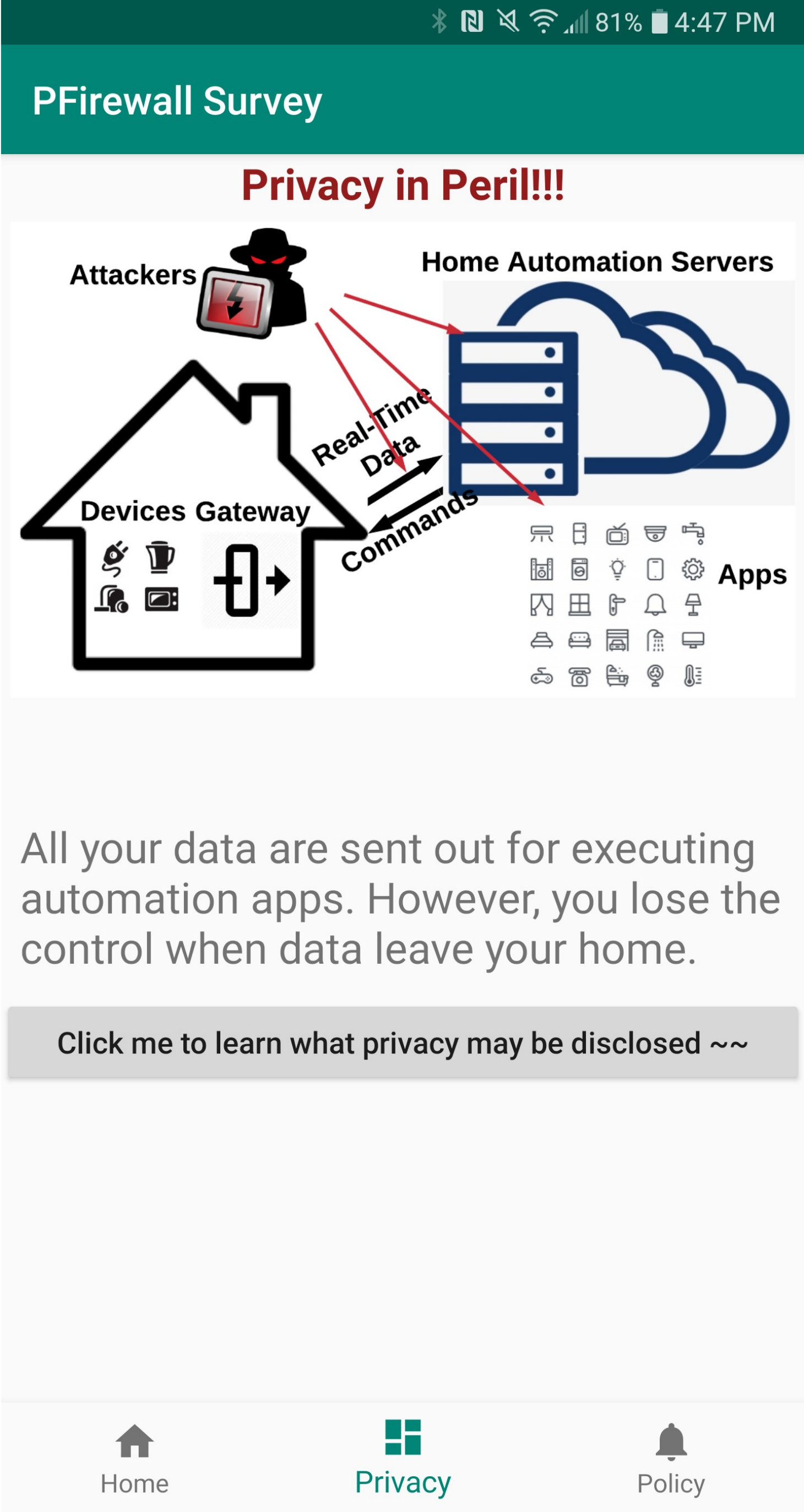}}
   \subfigure[]{
   \label{subfig_survey_privacy}
     \includegraphics[width=0.23\textwidth]{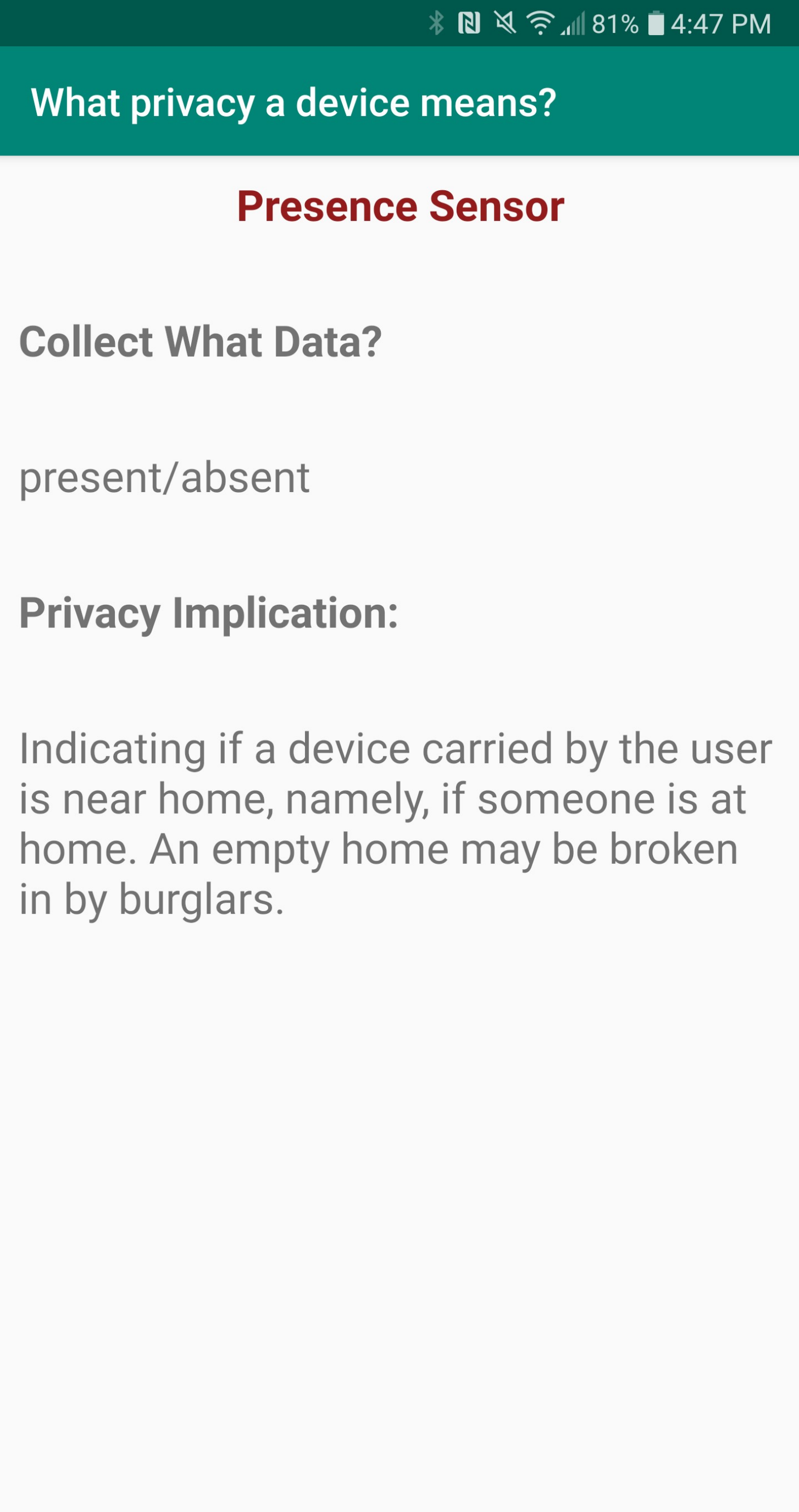}}
  \label{fig:survey}
  \caption{The PFirewall Survey mobile app used in the user survey.}
\end{figure*}

\subsection{Investigation on Popular Smart Home Platforms}
\label{section_investigation}
We study the privacy policies and practices on 7 popular cloud-based smart home platforms and 3 platforms that use other architectures for comparison. A brief summary is shown in Table~\ref{table_summary_privacy_policies}. ``Easy to access?'' shows if a privacy policy is explicitly displayed or prompted during the installation of the platform's products (especially apps). ``Collect device data?'' shows whether a privacy policy claims that the platform accesses users' devices during the services. ``Expose data to partners?'', ``Restrict data use on 3rd parties?'' and ``Privacy techniques'' show whether the platform claims to share users' data with third parties, whether it claims to restrict how third parties can legally use these data and what techniques it employs to protect user privacy during data sharing. ``Collect personal info.?'' shows whether a platform collects personally identifiable information from users during the registration process. ``Access device data?'' shows if the platform accesses device data while providing services. ``Expose data to partners?'' shows whether the platform provides device data to third-parties, including integrated third-party services. ``Access control before hub?'' and ``User controllable?'' indicate whether any access control mechanism is enforced before the platform's hub accesses device data and whether users can control the access between devices and the platform's hub.

\begin{table*}[ht] 
	\centering
	\caption{A summary of privacy policies and facts in some well-known platforms. AGG: aggregation; ANO: anonymization.}
	\newcommand{\tabincell}[2]{\begin{tabular}{@{}#1@{}}#2\end{tabular}}
	\renewcommand\arraystretch{1.1}
	\newcolumntype{P}[1]{>{\centering\arraybackslash}p{#1}}
	\newcolumntype{M}[1]{>{\centering\arraybackslash}m{#1}}
	\makeatletter
	\newcommand\notsotiny{\@setfontsize\notsotiny\@vipt\@viipt}
	\makeatother
	\notsotiny
	\begin{tabular}{c|ccccc|ccccc}
		\toprule
		\multirow{2}{*}{\tabincell{c}{\textbf{Platform}}} & \multicolumn{5}{c|}{\textbf{\tabincell{c}{Privacy Policy}}}  & \multicolumn{5}{c}{\textbf{Facts}}  \\ \cline{2-6} \cline{7-11}
		& \textbf{\tabincell{c}{Easy to\\ access?}} & \textbf{\tabincell{c}{Collect\\ device data?}} &  \textbf{\tabincell{c}{Expose data\\ to partners?}}  & \textbf{\tabincell{c}{Restrict data use\\ on 3rd parties?}} & \textbf{\tabincell{c}{Privacy\\ techniques}} & \textbf{\tabincell{c}{Collect\\ personal info.?}} & \textbf{\tabincell{c}{Access\\ device data?}} & \textbf{\tabincell{c}{Expose data\\ to partners?}} & \textbf{\tabincell{c}{Access control\\ before hub?}} & \textbf{\tabincell{c}{User \\controllable?}} \\
		\midrule
		Wink     & \CheckmarkBold & \CheckmarkBold    & \CheckmarkBold & \XSolidBrush & AGG &   \CheckmarkBold & \CheckmarkBold & \CheckmarkBold & \XSolidBrush & \XSolidBrush \\
		Iris     & \CheckmarkBold & \CheckmarkBold    & \CheckmarkBold & \XSolidBrush  & \XSolidBrush &    \CheckmarkBold &   \CheckmarkBold & \CheckmarkBold & \XSolidBrush & \XSolidBrush \\
		Vera     & \CheckmarkBold & \CheckmarkBold   & \CheckmarkBold & \CheckmarkBold   & AGG, ANO &   \CheckmarkBold &   \CheckmarkBold & \CheckmarkBold & \XSolidBrush & \XSolidBrush \\ 
		Lutron    & \CheckmarkBold & \XSolidBrush & \CheckmarkBold & \XSolidBrush  & \XSolidBrush  &   \CheckmarkBold &   \CheckmarkBold & \CheckmarkBold & \XSolidBrush & \XSolidBrush \\
		Thingsee   & \CheckmarkBold  &  \CheckmarkBold  & \CheckmarkBold & \CheckmarkBold  & AGG, ANO &   \CheckmarkBold &   \CheckmarkBold & \CheckmarkBold & \XSolidBrush & \XSolidBrush \\  
		SmartThings & \XSolidBrush  &   \CheckmarkBold    & \CheckmarkBold & \CheckmarkBold  & AGG, ANO  &  \CheckmarkBold & \CheckmarkBold  & \CheckmarkBold  & \XSolidBrush & \XSolidBrush \\
		EVRYTHNG   & \XSolidBrush  &   \XSolidBrush  &   \XSolidBrush & \XSolidBrush & \XSolidBrush  &   \CheckmarkBold &   \CheckmarkBold  & \CheckmarkBold  & \XSolidBrush & \XSolidBrush \\
		openHAB    & \XSolidBrush  &   \XSolidBrush  & \XSolidBrush  & \XSolidBrush & \XSolidBrush &   \CheckmarkBold &   \CheckmarkBold  & \CheckmarkBold & \XSolidBrush & \XSolidBrush \\
		Mozilla IoT   & \XSolidBrush  &   \XSolidBrush      & \XSolidBrush & \XSolidBrush & \XSolidBrush &   \CheckmarkBold &  \CheckmarkBold  & \XSolidBrush & \XSolidBrush & \XSolidBrush \\
		Apple HomeKit  & \CheckmarkBold  &   \XSolidBrush    & \XSolidBrush & \XSolidBrush &  \XSolidBrush   &   \CheckmarkBold &  \CheckmarkBold   & \XSolidBrush  & \XSolidBrush & \XSolidBrush \\
		\bottomrule
	\end{tabular}
	\label{table_summary_privacy_policies}
\end{table*}

Some privacy policies fail to increase user perceptions of sensitive data collection since they fail to 1) be easily accessible, or 2) use jargon-free words, or 3) claim sensitive data collection explicitly. Some policies, although claim sharing data with third-parties, do not claim any data protection techniques or any restriction policies to the third-parties. On the other hand, we found the fact that most of the studied platforms request personal-identifiable information from during registration, access sensitive data from IoT devices, and share data with business partners. However, most platforms do not have mechanisms to minimize the data access from users and do not provide interfaces to users for fine-grained controls on their sensitive data. Users are only capable of choosing whether to agree with the privacy policy. Once a device is connected to the platform, they cannot further decide how their deivces report data to the platforms.

\subsection{Time/Timer-related Automation}
\label{appendix_time_timer}
\begin{figure}[t]
	\centering
	\includegraphics[width=0.35\textwidth]{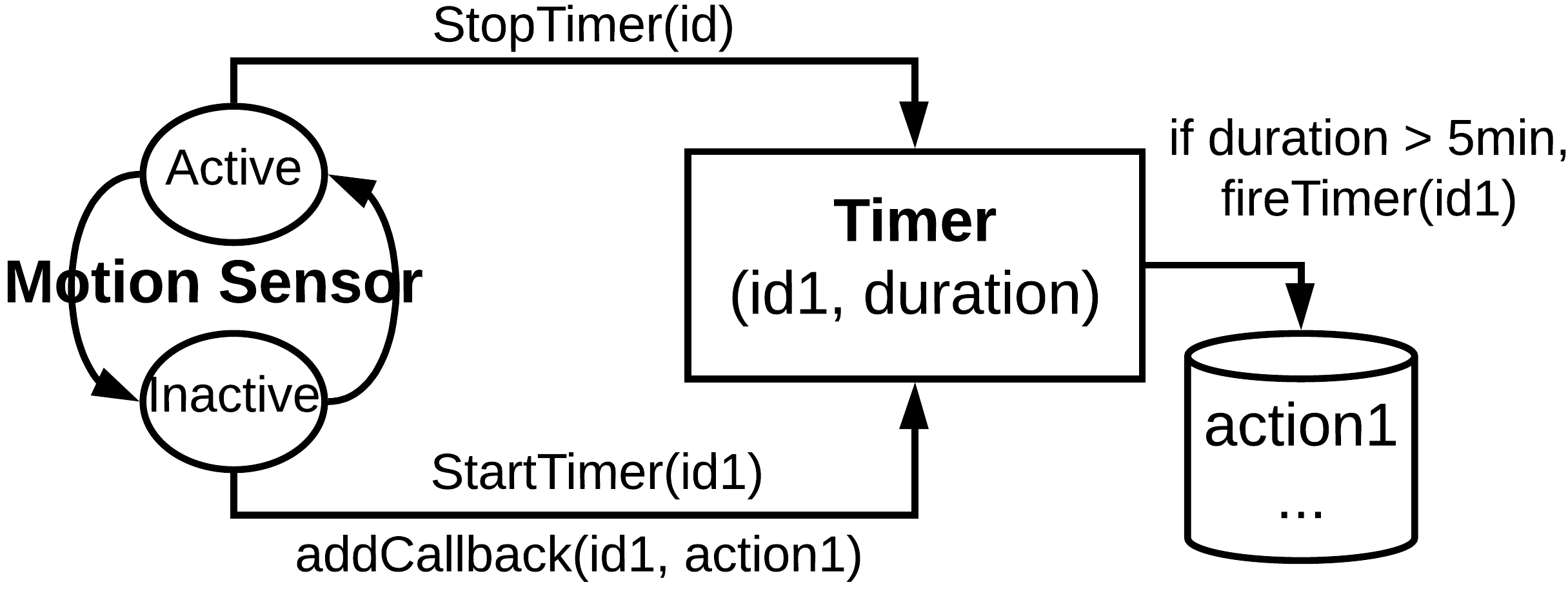}\\
	\caption{The workflow of how \tool handles a timer-related rule example. The methods are show in Table~\ref{table_methods_time_timer}. action1 is defined to report ``inactive'' to the platform with method \texttt{keep} and zero delay. Each timer maintains a list of actions which will be called when the timer's duration satisfies a certain constraint.}\label{fig_timer_example}
\end{figure}

\tool also deals with time-related automations. For instance, if a rule is defined as ``when the door is opened if time is after 18:00, turn on TV'', the derived policy needs to fetch system time for condition checking. When it comes to a timer-related automation, e.g., ``when motion sensor becomes inactive for 5 minutes, turn off the light'', multiple policies are bundled to operate by calling the methods for starting, stopping and firing a timer. Fig.~\ref{fig_timer_example} illustrates the workflow of how \tool handles this example.

\begin{table}
	\scriptsize
	\caption{Methods for dealing with timer-related automation}
	\renewcommand\arraystretch{1.1}
	\newcommand{\tabincell}[2]{\begin{tabular}{@{}#1@{}}#2\end{tabular}}
	\newcolumntype{P}[1]{>{\arraybackslash}p{#1}}
	\newcolumntype{M}[1]{>{\centering\arraybackslash}m{#1}}
	\begin{tabular}{P{2.5cm}P{5cm}}		
		\toprule
		\bf Method & \bf Description\\ \midrule
        \texttt{startTimer(id)} & Create or reset a timer with identity \texttt{id} \\
        \texttt{stopTimer(id)} & Stop and reset a timer with identity \texttt{id} \\
        \texttt{fireTimer(id)} & Fire a timer \texttt{id} and execute actions in its callbacks \\
        \texttt{addCallback(id,act)} & Add an action \texttt{act} to the callbacks of timer \texttt{id} \\
		\bottomrule 
	\end{tabular}
	\label{table_methods_time_timer}
\end{table}

\subsection{Interfacing with openHAB}
\label{appendix_openhab}

\subsubsection{Implementation}
\label{appendix_openhab_implementation}
\begin{figure}[tb]
	\centering
	\includegraphics[width=0.49\textwidth]{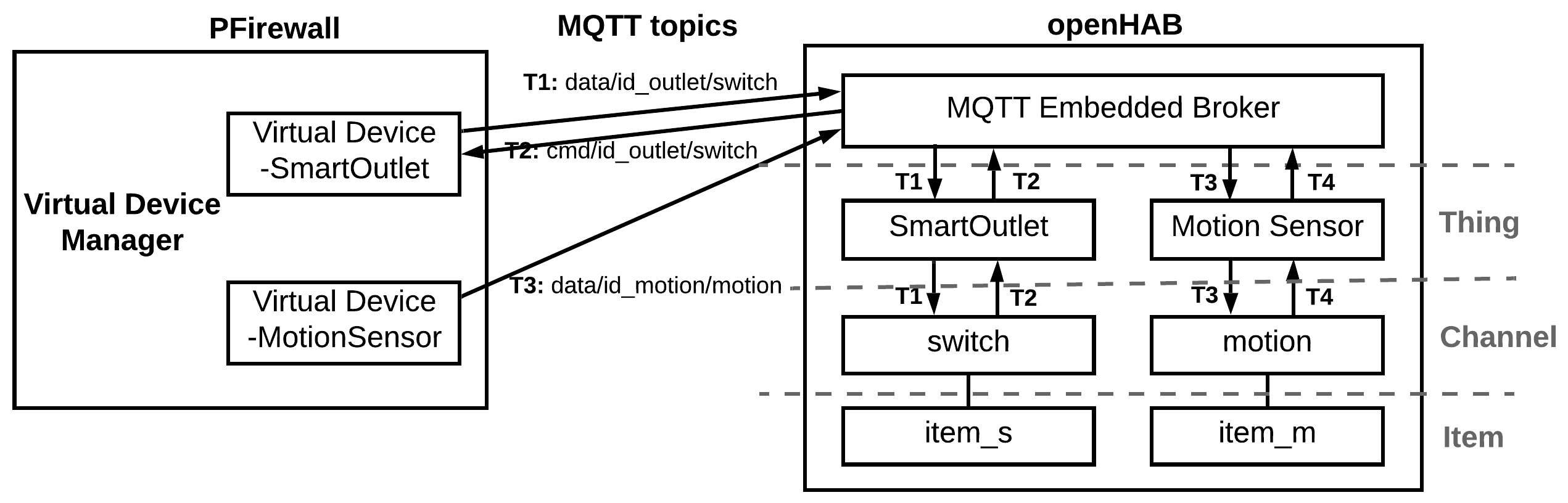}\\
	\caption{Overview of how the mediator interfacing with openHAB.}\label{fig_mqtt_implementation}
\end{figure}

We use the supported MQTT to interface with openHAB because it is a general connectivity protocol, allowing for virtualizing any device types with flexibility. Fig.~\ref{fig_mqtt_implementation} shows the high-level architecture of the integration. openHAB provides an embedded MQTT broker, so our work is to realize each virtual device (VD) as a MQTT client and create a Generic MQTT thing (supported by MQTT binding) in openHAB for the real device represented by the VD. A thing in openHAB has channels (equivalent to the concept ``attribute'' in SmartThings, e.g., motion, temperature, etc.) and each channel can be linked to an item (used for displaying values received by the linked channel and used as an interface for automation rules to interact with the real device). In openHAB, each MQTT thing channel can be configured as a MQTT client. By subscribing to the same MQTT topic (essentially a path-alike string), MQTT clients can publish/receive data to/from the topic. 

When a new device is added to \tool, a VD instance is created. If the real device is a sensor (e.g., motion sensor in Fig.~\ref{fig_mqtt_implementation}), the VD instance subscribes to a topic \texttt{data/{\{}device id{\}}/{\{}attribute{\}}} (e.g., \texttt{data/12345/motion}) for publishing data, where device id is generated randomly by \tool; if the real device is an actuator (e.g., smart outlet), the VD instance subscribes to a topic \texttt{data/{\{}device id{\}}/{\{}attribute{\}}} for publishing data and a topic \texttt{cmd/{\{}device id{\}}/{\{}attribute{\}}} for receiving commands. The MQTT bining in openHAB does not provide a device discovery function. To automatically add a thing and its channel in openHab, there are two choices: operating on the web interfaces or adding a configuration file in the \texttt{openhab/conf/things/} directory. We choose the latter approach to automate the process. By populating a string template with the same device id, attribute and topic information as the VD instance, \tool creates a MQTT thing by adding a thing file to the openHAB directory through a FTP service. Thus, the created MQTT thing can receive data from or send commands to the VD by subscribing to the same topics.    

\subsubsection{Evaluation}
\label{appendix_openhab_evaluation}
openHAB allows users to write automation apps with a domain specific language (DSL), which is adapted from Xbase \cite{Xtext2019}. However, openHAB does not provide official apps for installation. To test our openHAB integration, we develop 13 apps implementing the same rule semantics to work with the same devices, as shown in Table~\ref{table_rules}. We manually operate the real devices to trigger each rule for 20 times and find all apps are executed correctly.

\subsection{Complete Evaluation Result of Data Volume Reduction}
\label{appendix_complete_data_volume_comparison}
Due to page limits, we only present the result of one device for each device type in Table~\ref{table_reduction_result} in Section~\ref{section_data_reduction}. Table~\ref{table_reduction_result_complete} shows the complete list of all deployed devices in both testbeds.

\begin{table}[tb]
	\centering
	\scriptsize
	\caption{Comparison of reported data volume per device before and after the deployment of \tool. $VOL$: volume of reported data in {\tt SYS1} and {\tt SYS2}, respectively; $RR$: relative reduction rate. We present the result for each device type. See Appendix~\ref{appendix_complete_data_volume_comparison} for the complete result of all deployed devices.
	}
	\renewcommand\arraystretch{1}
	\newcommand{\tabincell}[2]{\begin{tabular}{@{}#1@{}}#2\end{tabular}}
	\newcolumntype{P}[1]{>{\raggedright\arraybackslash}p{#1}}
	\newcolumntype{M}[1]{>{\centering\arraybackslash}m{#1}}
	\newcolumntype{R}[1]{>{\raggedleft\arraybackslash}p{#1}}
	\begin{tabular}{M{0.4cm}M{1.05cm}M{1cm}M{0.4cm}M{1.05cm}M{1cm}M{0.4cm}}	
	\toprule
    \textbf{Dev} & \textbf{Attr} & $VOL$ & $RR$  & \textbf{Attr} & $VOL$       & $RR$ \\
    \midrule
    {\tt MU1}   &  contact  & 1924, 22  & 0.98  &  temperature  & 142, 6  & 0.96 \\ 
	{\tt MO1}   &  motion   & 2266, 47  & 0.98  &  temperature  & 307, 0  & 1 \\ 
	{\tt OL1}   &  switch   &  29, 0  & 1  &    &   &  \\
	{\tt OL2}   &  switch   &  19, 0  & 1  &    &   &  \\
	{\tt SL1}   &  switch   &  22, 0  & 1  &    &   &  \\
	{\tt SP1}   &  presence  & 34, 24  & 0.29  &    &   &  \\
	{\tt SP2}   &  presence  & 36, 1  & 0.97  &    &   &  \\
	{\tt SP3}   &  presence  & 30, 1  & 0.96  &    &   &  \\
	{\tt SP4}   &  presence  & 28, 1  & 0.96  &    &   &  \\
	{\tt SP5}   &  presence  & 26, 1  & 0.96  &    &   &  \\
	{\tt MU2}   &  contact  & 52, 24  & 0.54  &  temperature  &  118, 0 & 1 \\
	{\tt MU3}   &  contact  & 268, 58  & 0.78  &  temperature  & 131, 8  & 0.94 \\
	{\tt MU4}   &  contact  & 42, 42  & 0  &  temperature  &  109, 0 & 1 \\
	{\tt MO2}   &  motion  & 364, 68  & 0.81  &  temperature  &  173, 0 & 1 \\
	{\tt MO3}   &  motion  & 564, 21  & 0.96  &  temperature  &  157, 0 & 1 \\
	{\tt OL3}   &  switch  & 44, 0  & 1  &    &   &  \\
	{\tt OL4}   &  switch  & 49, 0  & 1  &    &   &  \\
	{\tt SL2}   &  switch  & 60, 0  & 1  &    &   &  \\
	{\tt SL3}   &  switch  & 68, 0  & 1  &    &   &  \\
	{\tt SL4}   &  switch  & 70, 0  & 1  &    &   &  \\
	{\tt SL5}   &  switch  & 42, 0  & 1  &    &   &  \\
	{\tt AM1}   &  motion  & 364, 0  & 1 
	\\
	{\tt AM1}   &  illuminance  & 1039, 1  & 0.99  &  humidity  & 668, 0  & 1 \\
	{\tt AM2}   &  motion  & 462, 0  & 1  
	\\
	{\tt AM2}   &  illuminance  & 1384, 0  & 1  &  humidity  & 893, 1  & 0.99 \\
	{\tt SP6}   &  presence  & 28, 12  & 0.57  &    &   & \\
	\bottomrule
	\end{tabular}
	\label{table_reduction_result_complete}
\end{table}

\subsection{User Study}
\label{appendix_user_study}
\subsubsection{Setup}
We conduct a user survey to study users' attitude and abilities towards defining customized data flow control policies with our policy templates (Section~\ref{section_policy_generation}). We recruit 20 adult participants who are knowledgeable about the concepts ``home automation'', ``smart home'' or ``IoT'' from our institutions. Participants completed the trial tasks of our ``PFirewall Survey'' app in our lab using smartphones we provided and after that answered several questions (see Section~\ref{appendix_example_questions}). 

We asked the participants to get familiar with a smart home setting where 10 automation rules (Fig.~\ref{subfig_survey_automation}) are configured to work with 15 devices (Fig.~\ref{subfig_survey_devices}). The app provides a page (Fig.~\ref{subfig_survey_tutorial}) to illustrate the architecture of the system and the potential risks of data leakage; we did not explain the content and ask questions about this page to avoid influencing the understanding of end-users by factors other than the interface itself. Besides, the app also provides an interface showing the list of 15 devices; when a device is selected, the app switches to a device detail page (e.g., Fig.~\ref{subfig_survey_privacy}) showing what data the device generates and what privacy risks are imposed if the data are leaked. In addition, policy templates (as shown in Fig.~\ref{fig_policy}) were provided for participants to define their own policies. After a 30-minute trial, participants were asked to answer questions.

\subsubsection{Results}
All 20 participants cared about their data privacy and thought it useful to define their own data flow policies for protecting privacy. However, 2 participants thought they would not spend time in defining policies even if an app is available. We collect the number of participants who had privacy concerns on each listed device. Cameras and smart speakers were the top two devices whose data are considered sensitive by the participants (19 and 16, respectively); half or more participants had concerns on the status data of smart locks, doors and windows (11, 13, 10, respectively); Each of humidity sensors, heaters, lights, powers and coffee makers is concerned by less than 3 participants. Except the listed devices, the participants also cared about the data privacy of smart TV, smart window blinds, smart outlet. 

Regarding the usability of our policy templates, 8 participants thought the templates are ``very easy'' to use and 12 participants thought them ``easy'' to use. 3 participants found that they cannot specify policies to control data by specifying multiple conditions with the templates, for example, the combination of an event and a specified time period. According to the feedback, we address this issue by allowing users to select another condition after a condition has been specified.

Overall, participants concern data privacy and hold a positive attitude in defining own policies with our templates. The result also shows that participants may overlook the privacy risks of some devices like humidity sensor and powers, which we have discussed in Section~\ref{privacy_gain}. Hence, data-minimization policies and user-specified policies could work together to achieve better privacy protection.

\subsubsection{Questions in the user study}
\label{appendix_example_questions}
\begin{enumerate}[leftmargin=*, itemsep=-1mm]
    \item Do you care about your data privacy if you use a smart home system?
    \\A. Yes
    \\B. No
    \item List the device(s) (from the given device list in our ``PFirewall Survey'' app) which you have privacy concerns if the device data are leaked.
    \item Do you think it is useful in general to control your own data to reduce privacy leakage risks?
    \\A. Yes
    \\B. No
    \item Would you spend time defining your own policies to control data if an app like ``PFirewall Survey'' is available for you to do so?
    \\A. Yes
    \\B. No
    \item Recall how our app guide you to define your own policies. Are the provided policy templates easy to understand and use?
    \\A. Easy
    \\B. Somewhat challenging but still able to use
    \\C. Not usable
    \item Do you find any policy that you think useful but the given templates fail to enable you to do so? If any, please list it.
\end{enumerate}

\subsection{IRB Approval}
\label{irb}
Our testbeds need to collect data from the testbed providers, including the 5 office members and 1 apartment member. Also, our user study involves 20 participants. We have received the approval from the IRB in the institution where all the above investigations are performed.

We value the participant privacy during our investigation processes. The data collected from both testbeds do not contain personally identifiable information and location data. The collected data will be transmitted to and stored in the password protected computer of one of the authors. Computers that store data have password-protected accounts and will be in a locked office that has limited access. Only the researchers identified on this protocol will have access to the data. Survey participants are asked to submit their questionnaire anonymously without revealing any personally identifiable information. The questionnaire will be stored in the locked office after analyses.

\end{document}